\documentclass[showpacs,aps,onecolumn]{revtex4}
\usepackage{epsfig}
\usepackage{graphicx}
\usepackage{amsmath,amssymb,amsfonts}
\usepackage{array}
\usepackage{url}
\usepackage{hyperref}
\usepackage{multirow}
\usepackage{float}
\usepackage{lineno}
\usepackage{xspace}
\usepackage{ulem}
\usepackage[usenames,dvipsnames]{color}
\usepackage{epstopdf}
\begin{document}
	
	\title{Laser-assisted (e,2e) study with twisted electron beam on H-atom}
	\author{Neha}
	\email{p20210062@pilani.bits-pilani.ac.in}
	
	\author{Nikita Dhankhar}
	\email{p20170427@pilani.bits-pilani.ac.in}
	
	\author{Raul Sheldon Pinto}
	\email{raul.sheldon@pilani.bits-pilani.ac.in}
	
	\author{Rakesh Choubisa}
	\email{rchoubisa@pilani.bits-pilani.ac.in }
	
	\affiliation{Department of Physics, Birla Institute of Technology and Science, Pilani, Pilani Campus Rajasthan, India}

	\begin{abstract}
		We study the laser-assisted twisted electron beam impact ionization of the hydrogen atom in coplanar asymmetric geometry. We develope the theoretical model in the first Born approximation. In the presence of the laser field, we treat the incident and scattered electrons as Volkov waves, the ejected electron, moving in the combined field of the laser and residual ion $H^+$, is described by a Coulomb-Volkov wave function . In this communication, we compare the angular profile of triple differential cross-section (TDCS) for laser-assisted twisted electron beam incidence with the plane-wave, laser-assisted plane wave, and field-free twisted electron beam for (e,2e) processes, for different orbital angular momentum (OAM) number ($m_l$) values. We analyze the influence of the laser parameters (photon exchanged, intensity ) on the angular distribution of the TDCS. We study the $(TDCS)_{av}$ for macroscopic target to examine the effect of opening angle $\theta_{p}$ of the twisted electron beam on the angular profile of TDCS. Our results clearly show the impact of laser parameters (electric field $\epsilon$ and number of photon exchanged ($l$)) and twisted electron beam parameters (OAM number ($m_l$) and opening angle ($\theta_{p}$)) on the angular distribution of TDCS.

	\end{abstract}
	\date{\today}
	\maketitle

  \section{Introduction}
  \label{intro}

	The studies of ionization processes on atomic, molecular and solid targets by impact of charged particles have seen significant growth over the decades. Electron impact ionization is a fundamental process that has numerous applications in various scientific and technological fields like mass spectrometry, plasma physics, Radiation Detectors (specifically in gas-filled detectors such as proportional counters and Geiger-Muller tubes), electron microscopy, and also in fields of chemistry and biology \cite{bartschat2016electron,shalenov2017scattering,de2019relativistic,CHAVEZ201989,BUG2017459}. 
	Particularly the electron-impact single ionization, also known as the (e,2e) process, has emerged as an important tool for analyzing the dynamics of ionization processes. In the (e,2e) process, an incident electron interacts with a microscopic target e.g. atom, and ejects one of the bound electrons of the target. The two outgoing electrons (the scattered and the ejected electrons) are detected coincidentally with their momenta fully resolved \cite{PhysRevA.97.062702}. The coincidence cross-sections for the (e,2e) process, here defined as Triple Differential Cross Sections (TDCS), gives the kinematically complete picture of the (e,2e) process \cite{bransden2003physics}. Since the first (e,2e) measurements
	reported independently in the late 1960s by Ehrhardt et al.\cite{Ehrhardt1969} and Amaldi et al. \cite{Amaldi1964}, many coincident experiments were performed on the atomic and molecular targets in various geometrical arrangements of the detected electrons \cite{walters1993directions,marchalant1997excitation,lahmam1991recent}. Various theoretical models of the (e, 2e) reactions on atomic and molecular targets were developed concurrently with these experiments. \\
	 The studies of (e,2e) processes help us understand the target's structure, electron correlations, and reaction mechanism during ionization process. The investigation of electronic collision processes in external fields, particularly in laser fields, has received much attention during the past few decades. The study of laser-assisted electron-atom collisions has played an important role in applied fields of physics such as plasma heating \cite{bartschat2016electron,christophorou2000electron}, semiconductor physics, gas breakdown, and fundamental atomic collision theory. Much useful information can be uncovered from the triple differential cross section (TDCS) measurements by using different geometrical arrangements. Various investigation in  \cite{mittleman2013introduction,francken1990theoretical,ehlotzky1998electron,ehlotzky2001atomic} provide an overview of laser-assisted atomic collisions. Initial studies neglected the target dressing effects. The unbounded electrons were described either as Volkov or Coulomb-Volkov states \cite{MOHAN1978399,cavaliere1980particle,PhysRevA.30.2759,PhysRevA.24.910,banerji1981electron,zangara1982influence,zarcone1983laser} Later, Joachain and colleagues investigated the cases of atomic hydrogen and helium \cite{joachain19882,martin1989electron}, considering the dressing effects of the initial and final (unbound) state, in the presence of laser field, and found that the differential cross section significantly varies. On this incorporation many authors have carryid out theoretical studies on laser-assisted the electron-impact ionization of atomic hydrogen, and single- and double-ionization of helium \cite{ghosh2009multiphoton,li2007ionization,li2005laser,chattopadhyay2005ionization,van2001electron,sanz1999semiclassical,makhoute1999light,taieb1991light,PhysRevA.56.4918}. Such investigations have also been expanded to include the bi-chromatic laser field  \cite{PhysRevA.56.3879,ghalim1999electron}. Atomic hydrogen and a few other inert gases have been the subject of experimental studies for the field-free scenario \cite{ehrhardt1986differential,PhysRevLett.22.89,PhysRevLett.48.1807}. The laser-assisted electron impact ionization experiment has been carried out in 2005 \cite{PhysRevLett.94.153201,hohr2007laser} on helium. This experiment paves a path toward related theoretical and experimental research.\\ 
	Significant and exciting breakthroughs have been made in creating electron vortex beams that carry orbital angular momentum (OAM) along the electron beam's propagation direction. It has been termed a ``twisted electron beam" \cite{uchida2010generation}. Unlike plane waves, TEBs possess non-zero OAM $(m_{l})$, which projects along the propagation direction of the beam. These beams carry a helical wavefront with phase  $e^{im\phi}$ having a well-defined azimuthal angle $\phi$ with propagation axis  \cite{lloyd2017electron,bliokh2017theory,larocque2018twisted}. Applications of vortex beams range from optical tweezers  \cite{he1995direct,o2002intrinsic} , microscopy \cite{furhapter2005spiral,baranek2013rotating}, micromanipulation \cite{galajda2002rotors}, astronomy \cite{berkhout2009using},and many more. The transfer of OAM to the target, in these cases, enables the manipulation of atoms and molecules in nanoscale dimension. Few theoretical studies exist for TEB collisions with atomic and molecular targets; such as the work conducted by Van Boxem, Partoens, and Verbeek \cite{van2014rutherford,van2015inelastic}. They have used the First Born Approximation (FBA) to study the influence of the OAM on the inelastic scattering processes of hydrogen-like targets. The same group has extended their study to examine the excitation of hydrogen by TEB projectiles, and they have derived a set of selection rules to show the OAM is directly transferred from the projectile to the target \cite{van2015inelastic}. Harris et al. (2019) \cite{harris2019ionization} report the TEB-induced ionization of hydrogen atoms by analyzing the full differential cross-section (FDCS) with a variety of factors of TEB. The recent study by Mandal et al. (2021) \cite{mandal2021semirelativistic} has shown the dependence of the Total Angular Momentum (TAM) number ($m_l$) and opening angle $\theta_p$ on the angular profile of the TDCS and the spin asymmetry for the relativistic electron impact ionization of the heavy atomic targets. Dhankhar et al. \cite{dhankhar2020double} investigated the double ionization of the He atom in the $\theta$-variable and constant $\theta_{12}$ modes for the incident twisted electron beam. Furthermore, Dhankhar et al. studied the effects of the twisted electron beam's OAM ($m_l$) and opening angle $\theta_p$ on the hydrogen molecule $H{_2}$ \cite{dhankhar2020electron}, water molecule $H{_2}O$  \cite{dhankhar2022triple} and also studied the dynamics of TEB on $CH_4$ and $NH_3$ molecules
	\cite{dhankhar2023dynamics}. In the recent year 2023, Harris et al \cite{harris2023controlling} studied the controlling electron projectile coherence effects using twisted electrons in the form of Laguerre-Gauss and Bessel electrons on  $H^+_{2}$ ion. 
	To the best of our knowledge, the laser-assisted electron impact ionization processes have only been studied for the plane wave incidence, and considerable work needs to be carried out for the laser-assisted TEB impact ionization. This paper aims to bridge this gap.\\
	The rest of the paper is structured as follows:
	Section  \ref{form} provides the theoretical treatment of laser-assisted (e, 2e) reactions and outlines the techniques employed to evaluate the first Born amplitudes for plane wave and the twisted
	electron beam.
	In section \ref{result}, we delve into the discussion of our numerical results, along with their corresponding physical implications.
	Section \ref{conc} encompasses our conclusions.
	Unless otherwise mentioned, atomic units are consistently used in the paper.

	\section{\textbf{Theory}}
	\label{form}
	
	\subsection{Plane Wave ionization cross-section}
	This section presents the theoretical formalism of a fast (e,2e) process in the presence of a laser field on hydrogen atoms by the plane wave and twisted electron beam incidence. The theoretical formalism of the (e,2e) process on the H-atom for a plane wave in the presence of a laser field and by TEB without a laser field is established. Research in the laser-assisted TEB field is not explored yet. In this work, we briefly describe the theoretical formalism for laser-assisted (e,2e) processes by TEB impact on H-atom. We choose the H-atom due to its relatively simple and its exactly known wave function. In addition, for simplicity, the laser field is treated classically, being linearly polarized, monochromatic, and spatially homogeneous over atomic dimensions. In the Coulomb gauge, we write the electric field $\epsilon(t)$ of the laser $\vec{\epsilon}(t)$ = $\vec{\epsilon_{0}}$ $sin($$\omega$t) the corresponding vector potential $\mathbf{A(t)}$ = $\mathbf{A_{0}}$ $cos$($\omega$t) where $\mathbf{A_{0}}$ = c$\mathbf{\epsilon_{0}}$/$\omega$ where $c$ is the velocity of light, $\epsilon_{0}$ is the laser-field strength and $\omega$ is the angular frequency of the laser field. The ionization electron-atom collision process, in the presence of a laser field with the transfer of $l$ photons, is described as:
	\begin{equation}
		e^-(\mathbf{k_i}) + H(1s) + l\omega  \rightarrow  H^+ + e^-(\mathbf{k_s}) + e^-(\mathbf{k_e})
	\end{equation}
	$\mathbf{k_i}$, $\mathbf{k_s}$, and $\mathbf{k_e}$ are the momenta of the incident, scattered, and ejected electrons respectively. For the (e,2e) kinematic arrangement, we select the Ehrhardt coplanar asymmetric geometry, which is such that fast-moving electrons of momenta 	$\mathbf{k_i}$ are incident on the target and fast-moving scattered electrons of momenta 	$\mathbf{k_s}$ are detected coincidentally with a slow-moving ejected electron of momenta 	$\mathbf{k_e}$. The three moments 	$\mathbf{k_i}$, $\mathbf{k_s}$, and $\mathbf{k_e}$ lie in the same plane. Moreover, the scattering angle $\theta_s$ of the fast electron is fixed and small, while the angle $\theta_e$ of the slow electron varies. We select the asymmetric Ehrhardt geometry because the theoretical and experimental results are available for the corresponding field-free (e,2e) case for H-atom. In this geometry, we write for the plane wave (e,2e) process,
	\begin{equation}
		e^- + H(1s) \rightarrow H^+ + e^-(	\mathbf{k_s}) + e^-(	\mathbf{k_e})
	\end{equation}
	As stated in the introduction, we are working in the FBA; thus, the central quantity to be evaluated is the first Born transition $(T-matrix)$ matrix element {\cite{joachain19882}} is:
	\begin{equation}
		T_{fi}^{B1}=-i\int_{-\infty}^{+\infty} dt\left< \chi _{k_{s}}\left ( \mathbf{r_{0}},t \right )\phi_{k_{e}}\left ( \mathbf{r_{1}},t \right ) |V| \chi _{k_{i}}\left ( \mathbf{r_{0}},t \right )\phi_{0}\left ( \mathbf{r_{1}},t \right )\right>
	\end{equation}
	\begin{equation}
		V(r) = \left|1/\mathbf{{r}_{01}}| - |{1/\mathbf{r_{0}}}|\right.
	\end{equation}
	
	$V(r)$ is the intreaction potential between the incident electron and atom. In equation (4), $\mathbf{r_0}$ denotes the coordinate of the incident (and scattered) electron, $\mathbf{r_1}$ is the
	coordinate of the target electron, and $|\mathbf{r_{01}}|$ = $\arrowvert$ $\mathbf{r_0}$-$\mathbf{r_1}$ $\arrowvert$ assuming the hydrogen is at the origin. The wave functions $\chi _{k_{i}}$ and  $\chi_{k_{s}}$ are the Volkov wave functions describing the motion of the incident and scattered electrons in the presence of the laser field, respectively. This can be described as {\cite{joachain19882}} :
	\begin{equation}
		\chi_{k_{i,s}}(\mathbf{r_0},t) = (2\pi)^{-3/2}exp[i(\mathbf{{k}_{i,s}}\cdot \mathbf{{r}_{0}} -\mathbf{{k}_{i,s}}\cdot {\mathbf{\alpha}_{0}} sin(\omega t) - E_{k_{i,s}} t)]
	\end{equation}
	where $E$ = $k^2/2$ and $\mathbf{\alpha_0}$ = ${\epsilon_{0}}/ \omega^2$, $\omega$ is the laser angular frequency. $\chi_{k_{i,s}}(\mathbf{r_0},t)$ is the exact solution of the time-dependent Schr$\ddot{o}$dinger equation for the laser-dressed `` free ''  particle.
	The wave functions
	$\phi_{0}\left ( \mathbf{r_{1}},t \right )$ and $\phi_{k_{e}}\left ( \mathbf{r_{1}},t \right )$ that appear in the equation (3) are the dressed states of the target atom in the presence of the laser field. Throughout our study, we have considered the electric field strength of the laser field to be much less than the atomic unit strength  ($\epsilon_{0}<< (e/a_{0}^{2}\simeq 5\times10^{11} V/m^{-1})$).The initial dressed bound state in the presence of external laser-field,  $\phi_{0}\left ( \mathbf{r_{1}},t \right )$  is obtained using the first-order time-dependent perturbation theory in the following manner,
	\cite{li1999laser}; 
	\begin{equation}
		\phi_{0}\left (\mathbf{r_{1}},t \right )=exp\left ( -iE_{0}t \right )exp\left ( -i\mathbf{a}\cdot \mathbf{r_{1}}\right )\left [ \psi_{0}\left ( \mathbf{r_{1}} \right )+\frac{i}{2}\sum_{n}\left [ \frac{exp\left ( i\omega t \right )}{E_{n}-E_{0}+\omega}-\frac{exp\left ( -i\omega t \right )}{E_{n}-E_{0}-\omega} \right ] M_{n0}\psi_{n}\left ( \mathbf{r_{1}} \right ) \right ] 
	\end{equation}.
	
	Where $\psi_{0}$ is the ground state wavefunction of a hydrogen atom,
	\begin{equation}
		\psi_{0} = \frac{1}{\surd\pi} \exp^{-r_{1}}
	\end{equation}
	with energy $E_{0}$ = -0.5 a.u, in the absence of an external field, $\mathbf{a}$ = $\mathbf{A/c}$ where $exp(-i\mathbf{a}\cdot \mathbf{r_{1}})$ is a gauge factor, ensuring the gauge consistency between the Volkov wave function (4) and the dressed target state wavefunction (5). Moreover we defined $ M_{n0}=\left< \psi_{n} \left|\vec{\epsilon_{0}}\cdot \mathbf{r} \right|\psi_{0}\right>$ as a dipole-coupling matrix element. For the spherically symmetric ground state $\psi_{0}$, the summation in Eq. (5) runs over the discrete and continuum hydrogen atom $p$ states.
	
	The dressed continuum state consisting of an ejected electron of momentum $\mathbf{k_{e}}$ moving under the combined influence of the residual ion and the laser field is proposed by \cite{joachain19882,PhysRevA.39.6178}:
	
	\begin{equation}
		\begin{aligned}
			\phi_{k_{e}}\left ( \mathbf{r_{1}},t \right )=exp\left ( -iE_{k_{e}} t \right )exp\left ( -i\mathbf{a}\cdot \mathbf{r_{1}} \right )exp\left ( -i\mathbf{k_{e}} \cdot \mathbf{\alpha_{0}} sin\omega t  \right ) \times \\ \left [ \psi_{C,k_{e}}^{-}\left ( \mathbf{r_{1}} \right )+ \frac{i}{2}\sum_{n}\left [ \frac{exp\left (i\omega t  \right )}{E_{n}-E_{k_{e}}+\omega}- \frac{exp\left (-i\omega t  \right )}{E_{n}-E_{k_{e}}-\omega}\right ]M_{n,k_{e}}\psi_{n}\left ( \mathbf{r_{1}} \right )+i\mathbf{k_{e}}\cdot \mathbf{\alpha _{0}} sin(\omega t) \psi_{C,k_{e}}^{-}\left ( \mathbf{r_{1}} \right )\right ];
		\end{aligned}
	\end{equation} 
	Where $\psi_{C,k_{e}}^{-}( r_{1})$ is the Coulomb wave function with incoming spherical wave behavior, corresponding to the momentum $k_{e}$ and normalized to a $\delta$ function in momentum space,
	\begin{equation}
		\psi_{c,k_{e}}^{-}=\left (2\pi   \right )^{-3/2}e^{\pi/2\mathbf{k_{e}}}e^{i\mathbf{k_e} \cdot \mathbf{r}} \Gamma\left ( 1+i/k_{e}\right )_1F_{1}\left [ -i/k_e,1,-i\left (k_{e}r_{1}+\mathbf{k_{e}}\cdot \mathbf{r} \right ) \right ],
	\end{equation}
	$\zeta$ = 1/$k_{e}$ is the Sommerfeld parameter, ${_{1}\textrm{F}_{1}(a,b,z)}$ is the conﬂuent hypergeometric function and $M_{n,k_{e}}=\left< \psi_{n} \left|\vec{\epsilon_{0}}\cdot \vec{r} \right|\psi_{C,k_{e}}^{-}\right>$.
	Substituting the expressions (5), (6), and (7) in the first Born T-matrix element and by integrating over time using the Fourier expansion of the $e^{(-i\mathbf{k_{e}}\cdot \mathbf{\alpha_{0}} sin(\omega t))}$, we obtain
	\begin{equation}
		T_{fi}^{B1}=\left ( 2\pi  \right )^{-1} i\sum_{l=-\infty}^{+\infty}\delta \left (E_{k_{s}} + E_{k_{e}} - E_{k_{i}}- E_{0}-l\omega  \right )f_{ion}^{B1,l}
	\end{equation}
	Here $f_{ion}^{B1,l}$ is the First Born scattering amplitude for the laser-assisted (e,2e) process, with the transfer of $l$ photons. The quantity is given by {\cite{joachain19882}};
	\begin{equation}
		f_{ion}^{B1,l}=f_I+f_{II}+f_{III}
	\end{equation}
	where, 
	\begin{equation}
		f_{I}=-2\mathbf{\Delta^{-2}} J_{l}\left ( \lambda  \right ) \left<\psi_{C,k_{e}}^{-}\left|exp\left ( i\mathbf{\Delta} .\mathbf{r} \right ) \right|\psi_{0} \right>,
	\end{equation}\\
	\begin{equation}
		f_{II}=i\mathbf{\Delta^{-2}}\sum_{n}\left< \psi_{C,k_{e}}^{-}\left| \exp(i\mathbf{\Delta} .\mathbf{r})\right|\psi_{n}\right>M_{n0}\left [ \frac{J_{l-1}\left ( \lambda  \right )}{E_{n}-E_{0}-\omega } -\frac{J_{l+1}\left ( \lambda  \right )}{E_{n}-E_{0}+\omega } \right ]
	\end{equation}\\
	\begin{equation}\label{12}
		\begin{aligned}
			f_{III}=i\mathbf{\Delta^{-2}}\sum_{n}\left< \psi_{n}\left| \exp\left ( i\mathbf{\Delta} .\mathbf{r} \right )\right|\psi_{0}\right>M_{n,k_{e}}^{*}\left [ \frac{J_{l-1}(\lambda )}{E_{n}-E_{k_{e}}+\omega }-\frac{J_{l+1}\left ( \lambda  \right )}{E_{n}-E_{k_{e}}-\omega } \right ]\\
			-\mathbf{\Delta^{-2}}\mathbf{k_{e}}.\mathbf{\alpha_0}\left [J_{l-1} \left (\lambda   \right )-J_{l+1}\left ( \lambda  \right )  \right ]\left< \psi_{C,k_{e}}^{-}\left| exp\left ( i\mathbf{\Delta} .\mathbf{r} \right )\right|\psi_{0}\right>
		\end{aligned}
	\end{equation}
	
	In these equations, $J_{l}$ is the Bessel function of order $l$, the quantity $\lambda$ = $(\mathbf{\Delta}- \mathbf{k_{e}})\cdot \mathbf{\alpha_{0}}$ where $\mathbf{\Delta}$ = ($\mathbf{k_{i}} - \mathbf{k_{s}}$) is the momentum transfer in the collision.
	The triple differential cross-section for the Laser-assisted (e,2e) process with the transfer of $l$ photons, in First Born Approximation, is given by,
	\begin{equation}
		\frac{d^{3}\sigma_{ion}^{B1,l}}{d\Omega _{s}\Omega_{e}dE}= \frac{k_{s}k_{e}}{k_{i}}\left|f_{ion}^{B1,l} \right|^{2}.
	\end{equation}
	We emphasize that the amplitude $f_{1}$ is consistent with the first Born treatment, ignoring the target dressing effects \cite{zarcone1983laser}.
	In this approximation, the TDCS reduces to 
	\begin{equation}
		\frac{d^{3}\sigma_{ion}^{B1,l}}{d\Omega _{s}\Omega_{e}dE}= \frac{k_{s}k_{e}}{k_{i}}\left|f_{1} \right|^{2}.
	\end{equation}
	with 
	\begin{equation}
		f_{I}=-2\mathbf{\Delta^{-2}} \left<\psi_{C,k_{e}}^{-}\left|exp\left ( i\mathbf{\Delta} \cdot \mathbf{r} \right ) \right|\psi_{0} \right>,
	\end{equation}
	
	\subsection{Twisted beam ionization}
	\label{TEB}
	Having described the theory of the laser assisted (e,2e) process by plane wave electron beam, we now do the same for the twisted electron beam. The formalism of the (e,2e) process by TEB is the same as for the plane wave as described in section A for H-atom, except that a twisted electron beam replaces the incident plane wave. The twisted electron beam is a superposition of plane and carries an orbital angular momentum (OAM) $m_l\hslash$ along the propagation direction. The momentum vector $\mathbf{k_i}$ of the incident twisted electron beam is described as  \cite{dhankhar2020double}
	\begin{equation}
		\mathbf{k}_i = (k_i\sin{\theta_p}\cos{\phi_p})\hat{x}+(k_i\sin{\theta_p}\sin{\phi_p})\hat{y}+(k_i\cos{\theta_p})\hat{z}.  
	\end{equation}
	where $\theta_{p}$ and $\phi_{p}$ are the polar and azimuthal angles of the $\mathbf{k_i}$, with the beam propagation direction along the z-axis. Here the polar angle is also known as the opening angle $\theta_p$ = $\tan^{-1}\frac{k_{i\perp}}{k_{iz}}$, defined as the angle that momentum vector makes with the z-axis. The components ${k_{i\perp}}$ and ${k_{iz}}$ are the perpendicular and the longitudinal components of the incident momentum $\mathbf{k_{i}}$, respectively.
	Since  analytically it is difficult to achieve exact alignment of the projectile with the target, so one needs the more generalized form of Bessel beam \cite{PhysRevA.92.012705}, such that
	\begin{equation}
		\psi_{\varkappa m}^{\left ( tw \right )}\left ( \mathbf{r_{0}} \right )= \int_{0}^{\infty }\frac{dk_{i\perp}}{2\pi }k_{i\perp}\int_{0}^{2\pi}\frac{d\phi_p}{2\pi}a_{\chi m}\left ( k_{i\perp} \right )e^{i \bf{k_{i}.r_0}}e^{-i\bf {k_{i}}.b}
	\end{equation}
	Where $\textbf{b}$ is a vector that indicates the extent of the transverse orientation of the incident twisted electron beam with respect to the incident beam direction. $\textbf{b}$ is also known as the impact parameter, defined as $\textbf{b}$ = $b$ cos$\phi_{p}$ $\hat{x}$ + $b$ sin$\phi_{p}$ $\hat{y}$ , $b$ is the magnitude and $\phi_{p}$ is the azimuthal angle of $\textbf{b}$.
	
	We use the Bessel beam instead of the plane wave to calculate the transition matrix element for the twisted electron beam (19). Thus we get the transition matrix of TEB $T_{fi}^{tw}(\varkappa, \mathbf{\Delta})$ in terms of the plane wave beam transition amplitude $T_{fi}^{pw}(\mathbf{\Delta})$ in equation(3) for given $\mathbf{\Delta}$ = $\mathbf{k_{i}}$ - $\mathbf{k_{s}}$ \cite{dhankhar2020double} as:
	\begin{equation}
		T_{fi}^{tw}\left ( \varkappa,{\mathbf{\Delta}},\mathbf{b} \right )= \left ( -i \right )^{m}\sqrt{\frac{\varkappa}{2\pi}}\int_{0}^{2\pi}\frac{d\phi_{p}}{2\pi}e^{im\phi_{p}-ik_{i\perp b}}T_{fi}\left ( \mathbf{\Delta} \right ),
	\end{equation}
	where $k_{i\perp} \cdot b $ = $\mathbf{\varkappa}$ $b$ $cos(\phi_{p} - \phi_{b})$ and $\phi_{b}$ is the azimuthal angle of the impact parameter.
	The magnitude of the momentum transfer from the incident twisted electron beam to the target atom is given by, 
	\begin{equation}
		\Delta^{2} = k_{i}^2 + k_{s}^2 - 2 k_{i} k_{s} cos(\theta),
	\end{equation}
	where,
	\begin{equation} 
		\cos(\theta) = \cos(\theta_p)\cos(\theta_s) + \sin(\theta_p)\sin(\theta_s)\cos(\phi_p-\phi_s).
	\end{equation}
	Here $\theta_{s}$ and $\phi_{s}$ are the polar and azimuthal angles of the momentum $\mathbf{k_{s}}$.

	In this communication, the target is assumed to be located along the incidence direction of the twisted electron beam, which is the z-axis. Thus by substituting $\textbf{b}$ = 0 in equation (19), the transition amplitude for the twisted electron beam can be rewritten as 
	\begin{equation}
		T_{fi}^{tw}\left ( \varkappa,{\Delta},\mathbf{b} \right )= \left ( -i \right )^{m}\sqrt{\frac{\varkappa}{2\pi}}\int_{0}^{2\pi}\frac{d\phi_{p}}{2\pi}e^{im\phi_{p}}T_{fi}\left ( {\bf{\Delta}} \right )
	\end{equation}
	For the computation of the TDCS for the TEB, in the considered coplanar asymmetric geometry for which $\phi_{s}$ = 0, we integrate the equation (22) over $\phi_{p}$.
	The case where a single atom scatters the projectile is hard to achieve experimentally. Thus, it is vitally important to consider a macroscopic target.
	The cross-section for such targets is calculated by taking the average of the plane wave matrix element over all possible impact parameters, \textbf{b}. The average cross-section $(TDCS)_{av}$ in terms of plane wave cross-section can be written as \cite{harris2019ionization} :
	\begin{equation} \label{15}
		(TDCS)_{av}=\frac{1}{2\pi\cos\theta_p}\int^{2\pi}_{0}d\phi_p \frac{d^3\sigma(\mathbf{\Delta})}{d\Omega_{e}d\Omega_{s}dE_{e}}.  
	\end{equation}
	In equation (24), $\frac{d^3\sigma(\mathbf{\Delta})}{d\Omega_{e}d\Omega_{s}dE_{e}}$ is similar to the plane wave TDCS depending on momentum transfer. From equation (24), one can see that TDCS averaged over impact parameter \textbf{b} is independent on the OAM $(m_{l})$ of the incident twisted electron beam.


	


	%

	\section{Results and Discussions}
	\label{result}
	\subsection{Plane wave vases LA plane wave and LA TEB}

	In this section, we present the result of our calculations and their inferences for the Laser-assisted twisted electron beam impact ionization of hydrogen in coplanar asymmetric geometry using First Born approximation. We have benchmarked our results for the interaction of the plane electron wave in the absence and presence of the laser field. These are in accordance with the published results. Precisely, we have replicated our results for the impact of the twisted electron beam, which matches well with Harris et al.\cite{harris2019ionization}. The parameters used impact energy $E_i$ = 0.5 keV and $E_e$ as 20 eV, 50 eV, and 100 eV for $\theta_{s}$ = 100 mrad with various $\theta_{p}$ values, i.e., 1 mrad, 10 mrad, 100 mrad and $m_{l}$ = 1 and 2. We notice the magnitude of TDCS contrasts with the published results by Harris et al.\cite{harris2019ionization}. We suspect that this contrast is due to the use of a Coulomb wave in our calculations rather than a plane wave as an ejected electron beam.
	In our work, we have used a geometrical arrangement in which the electric field vector of polarization, $\hat{\mathcal{E}}$ is parallel to the incident momentum $ \mathbf {\hat{k}}_{i}$. The electric field strength is kept fixed at $\epsilon_{o}$  = $10^7$ V/cm and corresponds to the laser intensity 1.32 × $10^{11}$ W $cm^{-2}$. We analyze the TDCS as a function of the emission angle of the ejected electrons for the fixed number of photons $l$ exchanged in the collisions and fix scattering angle $\theta_s$ (angle of the fast scattered electron). We study the effects of the different parameters of the laser beam and the twisted beam on the angular profile of the (e,2e) cross-sections.
	Further, we present the angular profile of the TDCS for the ejected electron energies $E_e$= 20eV, 30eV, 40 eV, and 50eV and compare our results of the plane wave without laser field (\textbf{PW}) with the laser-assisted plane wave (\textbf{LA-PW}) and laser-assisted twisted electron beam (\textbf{LA-TEB}) for different values of the orbital angular momentum $m_{l}$ = 1, 2, 3 and 4 (where $\mathbf{b}$ = 0 throughout the calculations).

	\begin{figure*}[htp!]
		\begin{tabular}{ccc}
			\includegraphics[width=6.00cm]{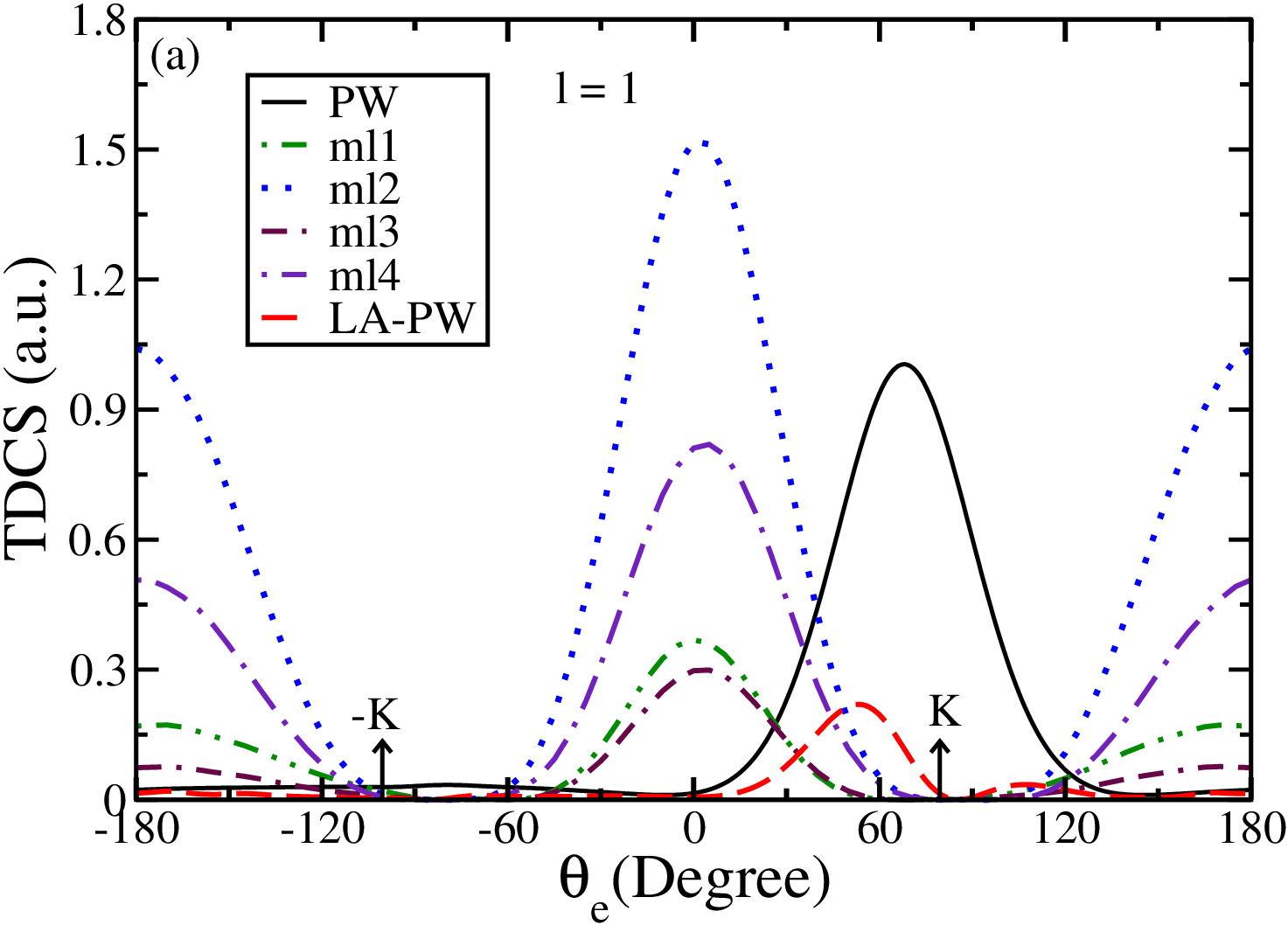}\ &
			\includegraphics[width=6.00cm]{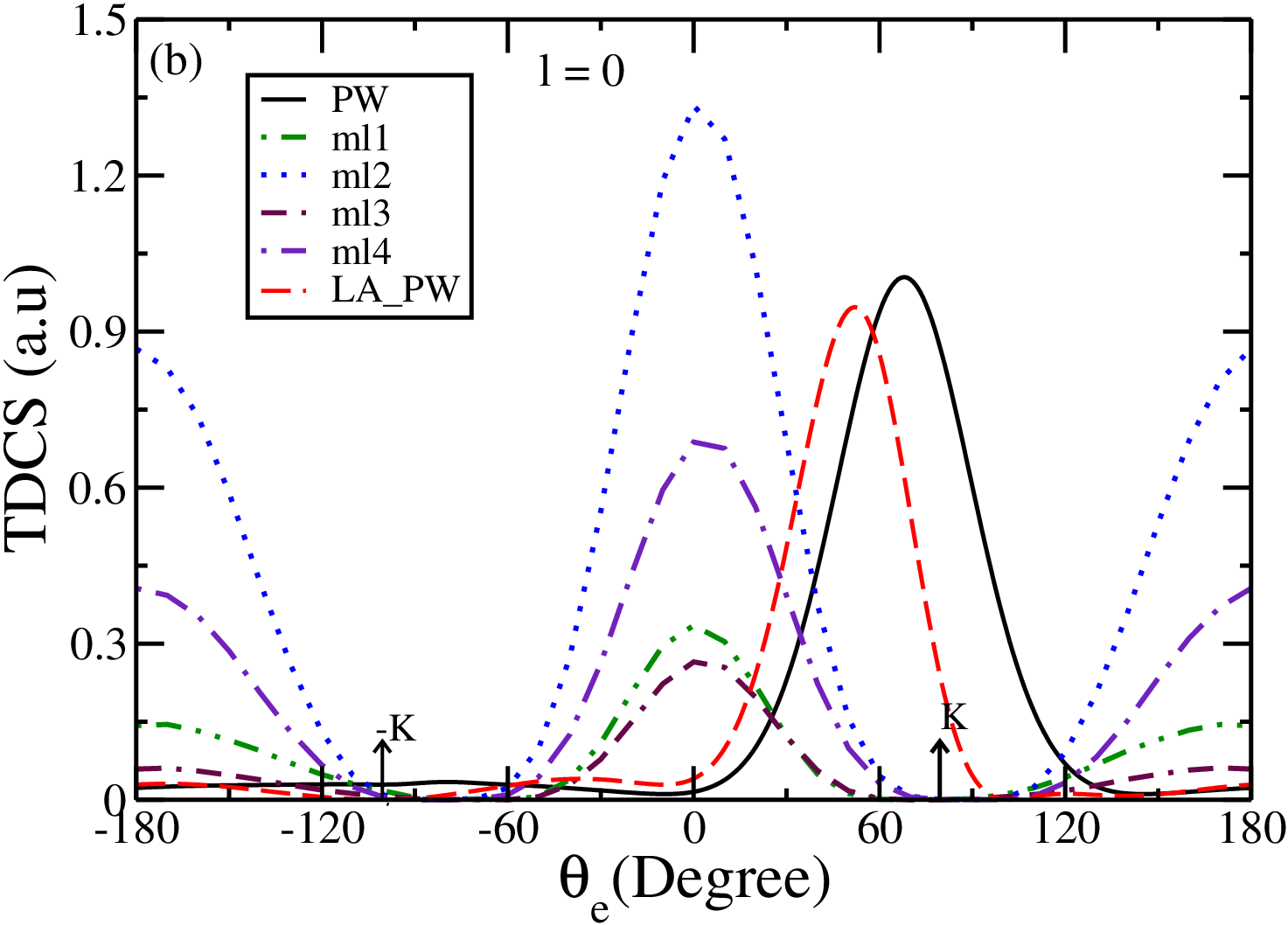}\
			&\includegraphics[width=6.00cm]{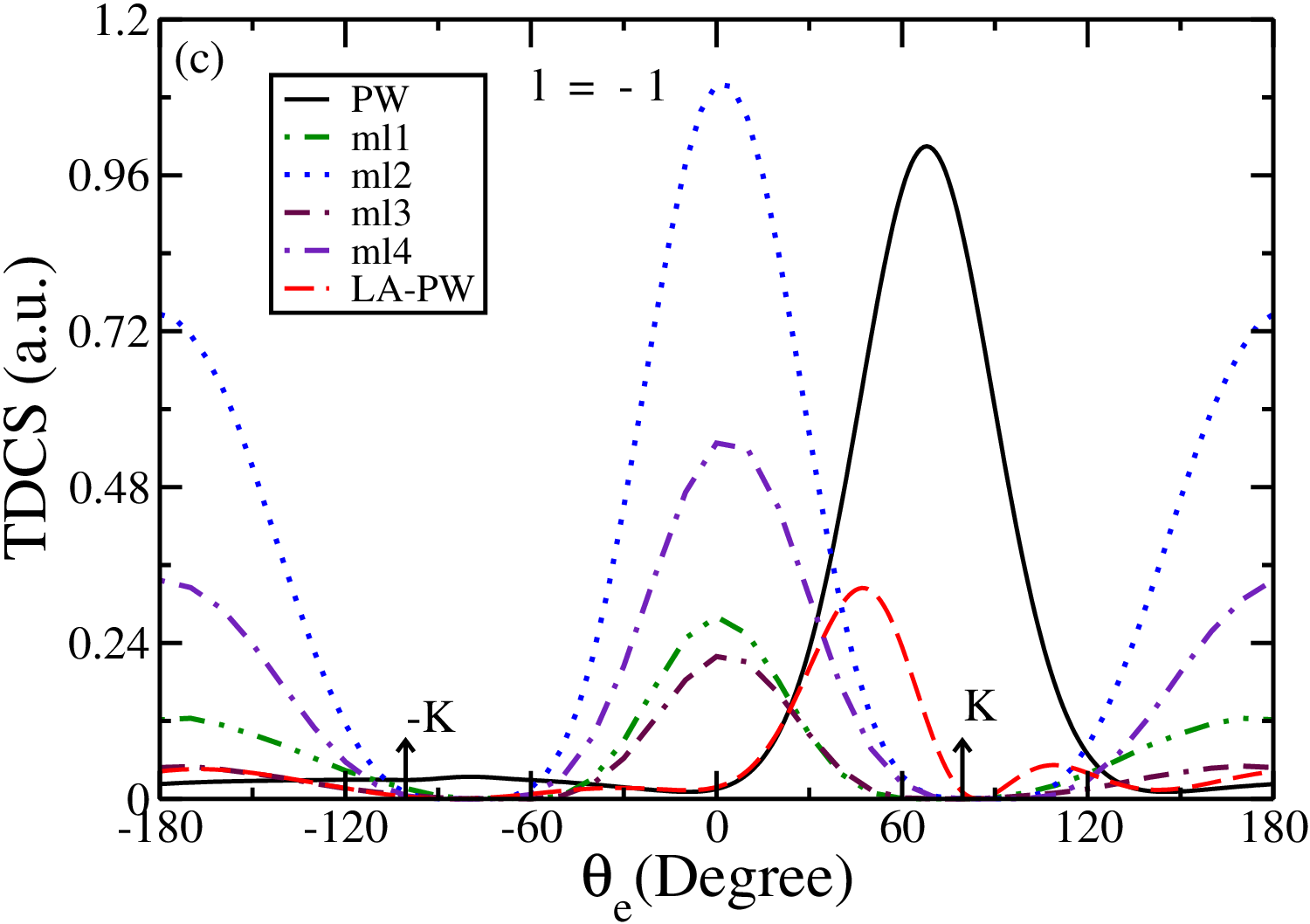}
		\end{tabular}
		\caption{TDCS as a function of the ejected electron angle $\theta_e$ for LA-TEB for (e,2e) process in asymmetric coplaner geometry for $\epsilon_0$ $\parallel$ $k_i$. The kinematics used here is $E_i$ = 500eV, $E_e$ = 20 eV, the laser amplitude $\epsilon_{0}$ = $10^7$V/cm and the frequency $\hbar$ $\omega$ = 1.17 eV (Nd: YAG laser).Keeping scattering angle ($\theta_{s}$) = Opening angle ($\theta_{p}$) = 100mr. TDCS is plotted for different values of $m_l$ and $l$. The TDCS magnitude of LA-PW for $l$ = $\pm$1 is normalised by a factor of 100 and for $l$ = 0 normalised by a factor of 500, and LA-TEB results are normalized by a factor of 100 to compare with the PW results.} 
		\label{fig:1}  
	\end{figure*}
	In Fig 1-8, we have represented PW, LA-PW, and LA-TEB with $m_l$ = 1, 2, 3, and 4 with solid, dashed, dotted-dotted-dashed, dotted, dashed-dashed dotted, and dotted dashed curves respectively .\\
	In figure \ref{fig:1}, we plot the triple differential cross section corresponding for the ionization of hydrogen atoms from the ground state. We begin with the analysis of the TDCS for plane wave, then plane wave in the presence of a laser source and a twisted electron beam along with a laser field. The incident and ejected electron energies used are $E_{i}$ = 500eV and $E_{e}$ = 20 eV, respectively. For the TEB, we keep the $\theta_{s}$ = $\theta_{p}$ for all values of $m_{l}$ from 1 to 4 with a unity difference. 
	To effectively compare the TDCS of LA-TEB with that of plane wave and LA-plane wave, it is important to acknowledge the established characteristics and trends associated with plane wave and LA-plane wave. The plane wave predicts the binary and shallow recoil peaks in the direction momentum transfer and opposite to it respectvely. As the energy of ejected electron increases, the absolute magnitude of the binary peak for a plane wave decreases while the magnitude of the recoil peak increases (refer to Figure 1-8, black solid curve). Additionally, as the energy of the ejected electron increases, the direction of linear momentum transfer shifts towards the incident beam. The LA plane wave scattering amplitude is calculated using equation (11) and complete TDCS with the help of equation (15). Similar to the plane wave, the LA plane wave also exhibits a dual-peak structure, binary peak and recoil peak. The electron-electron interaction is responsible for the binary peak, whereas the attraction between the electron and the nucleus governs the recoil peak. When considering LA-plane waves, it is noticeable that  TDCS magnitude decreases, and both the binary and recoil peaks exhibit a shift in comparison to the plane wave (see red dashed curve). On absorption ($l$ = 1)(see red dashed curve \ref{fig:1}(a)) and emission of one photon ($l$ = -1) (see red dashed curve \ref{fig:1}(c)), the symmetry of binary peaks vanishes, and we see smaller lobes accounting for the laser field for the LA plane wave. When examining the LA-TEB, like the plane wave, the TDCS of LA-TEB also shows two peak structures, forward peak, and backward peak. We observed that peaks for LA-TEB are no longer aligned in the direction of plane wave linear momentum transfer. As the TEB being the superposition of tilted plane waves, no single momentum
	transfer can be deﬁned,  but rather a ring of momentum transfer vectors exists. As the incident momentum now possesses a component perpendicular to the original plane wave's direction, the orientation of the 'momentum transfer' direction will undergo a deviation from its original plane wave configuration. The magnitude of LA-TEB TDCS is also smaller than that of plane wave TDCS, similar to LA plane wave TDCS. Like the plane wave, the TDCS magnitude decreases as the ejected electron energy increases. For LA-TEB, the angular profile of TDCS exhibits symmetry relative to the incidence direction (0$^\circ$). The angular distribution of TDCS obtained in the case of the LA-TEB with odd OAM $(m_l)$ values differ significantly from those of the even OAM $(m_l)$ values. Also, the magnitudes of TDCS for even values of OAM are greater than those of the odd OAM values. The magnitudes of the TDCS are seen to follow the descending order for $m_l$ = 2, 4, 1, 3. We also notice a slight shift in the forward peak in the clockwise direction on the increase of $m_l$. This clearly shows the influence of the OAM $(m_l)$ on the angular profile of TDCS. Upon taking into account the influence of exchanged photons between the laser field and atom-electron system, we found the angular profile of TDCS for emission -$l$ and absorption $l$ of a  photon is similar while the magnitude is different since $j_{-l}(\lambda)$ = $(-1)^l J_l(\lambda)$. On comparing the outcomes of the exchanged photon effects on the TDCS for the LA plane wave and LA-TEB, we noticed that for the LA plane wave, the angular profiles of TDCS for absorption ($l$ = 1)(see fig 1 (a)) and emission ($l$ = -1) (see fig 1 (c)) are different than those for no net exchange of photons ($l$ = 0) (see fig 1 (b)). But for  LA-TEB, we observe minimal disparities in the angular profiles of TDCS for $l$ = 0 and $l = \pm1$, except for the decrease in TDCS magnitude corresponding to a reduction in the number of photons. (Note that we have scaled for $l$ = 1, 0, and -1 in Fig 1).
	\begin{figure*}[htp!]
		\begin{tabular}{ccc}
			\includegraphics[width=6.00cm]{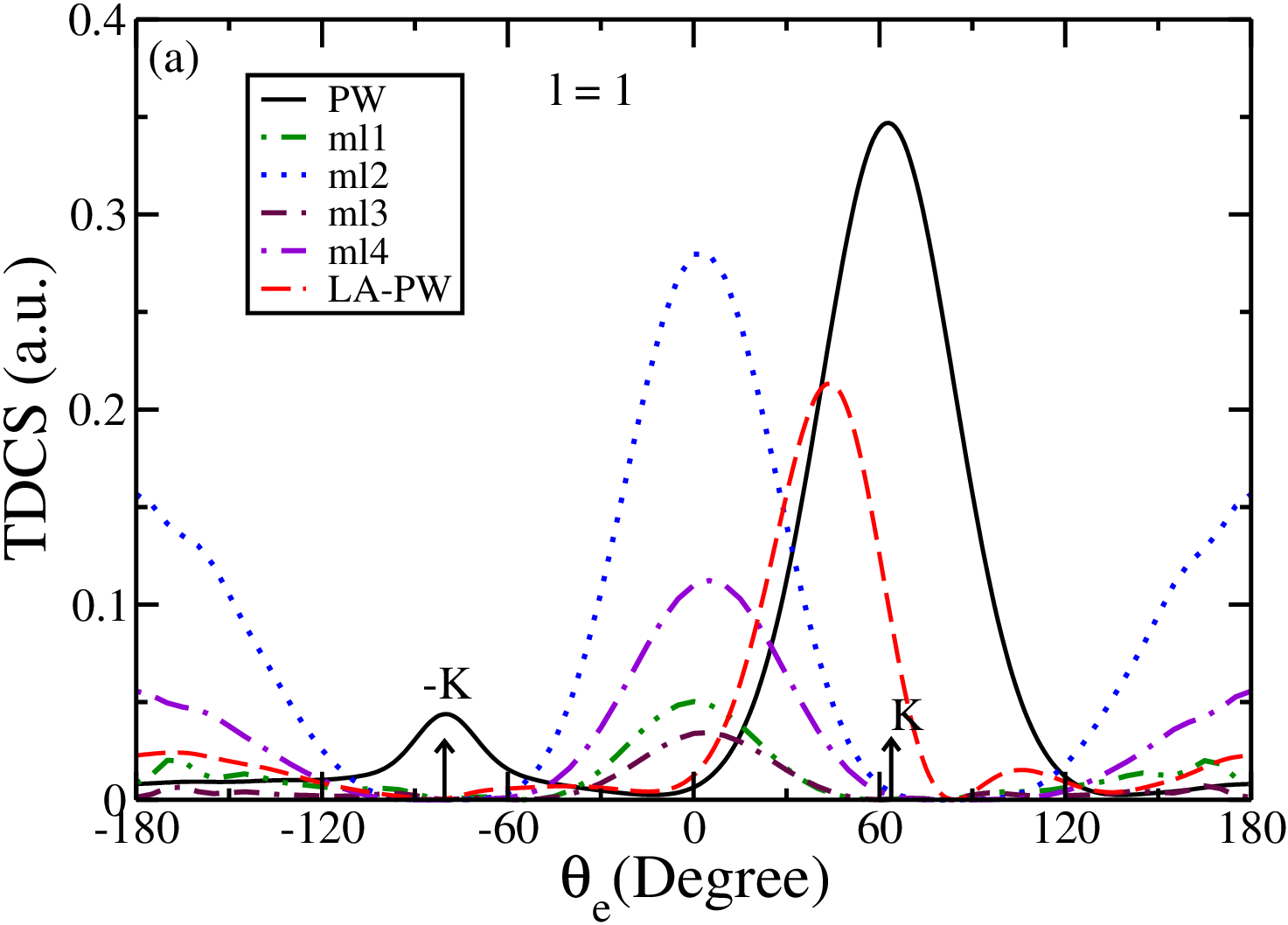}\ & \includegraphics[width=6.00cm]{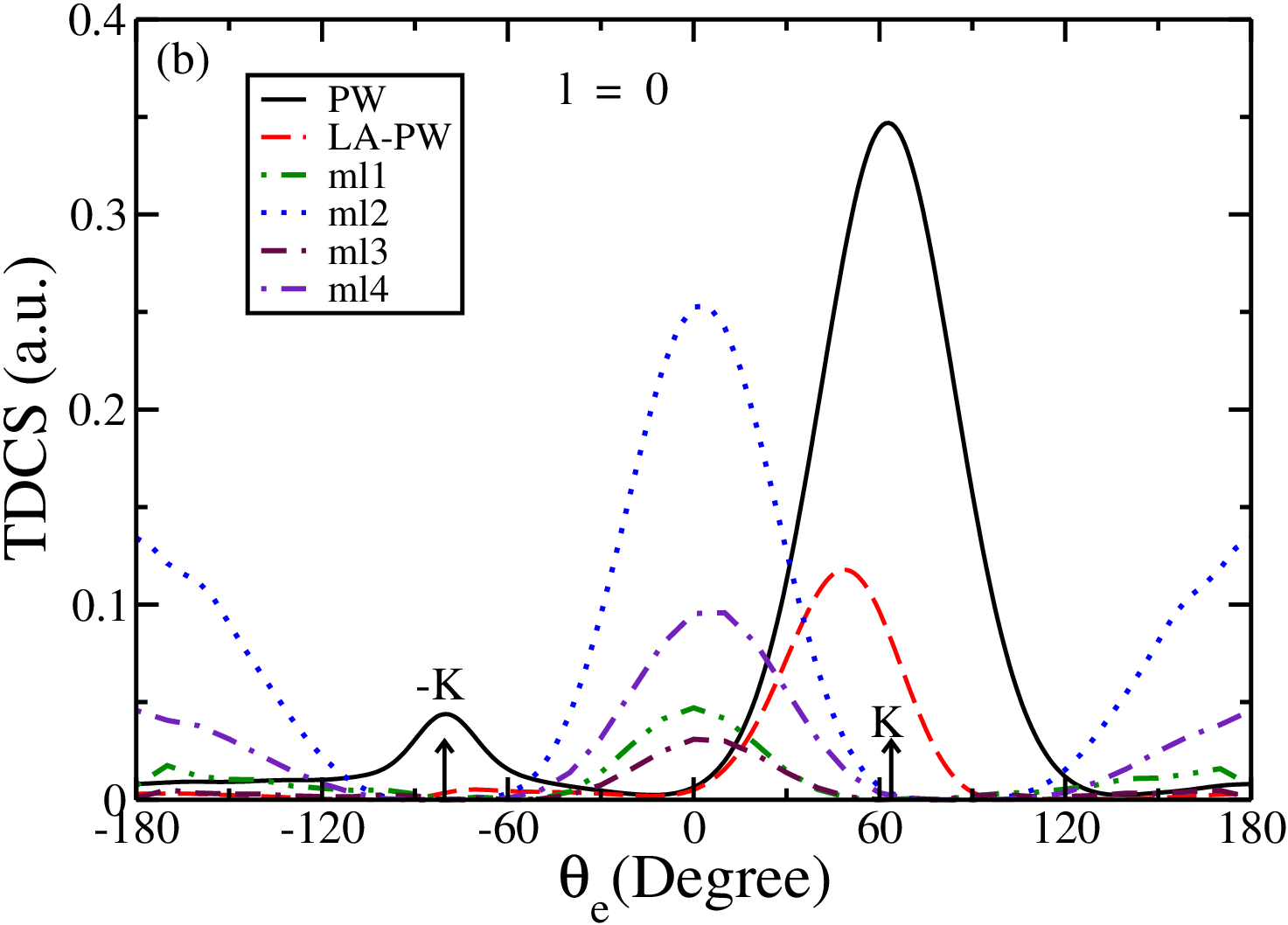}
			& \includegraphics[width=6.00cm]{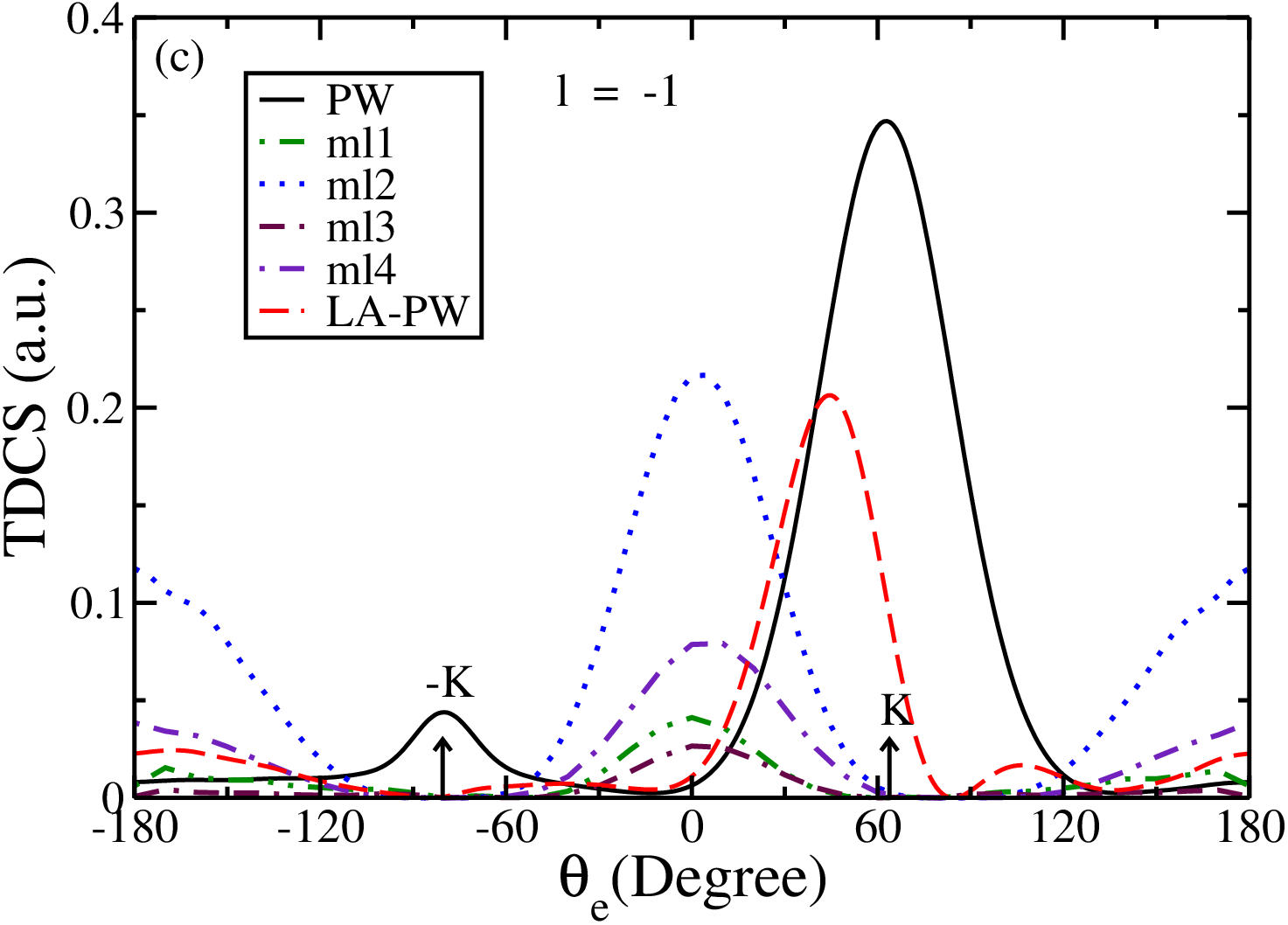}
		\end{tabular}
		\caption{Same as the figure 1 except the kinematics $E_e$ = 30eV. The magnitude of TDCS for LA-PW and LA-TEB is normalized by a factor of 100 to compare with the PW results.}   
		\label{fig:2}
	\end{figure*}
	In the figure \ref{fig:2} we plot TDCS for the ejected electron energy 30eV. We found that, for PW as the ejected energy increases, the absolute magnitude of the TDCS decreases while the recoil peak's magnitude increases. In the case of a plane wave (PW), as the ejected energy increases, the binary peak and the recoil peak shift toward the beam direction. For the LA-PW, we observe a shifted binary peak at a smaller angle compared to the PW (red dashed line in Figure 2), along with different smaller lobes. Notably, the recoil peak disappears in LA-PW calculations. Regarding the LA-TEB, the impact of increased ejected energy on the different values of the OAM can be seen in the angular distribution of TDCS, as it is different for different values of $m_l$. The peaks near $\theta_e = \pm180^\circ$ become distorted for odd OAM values; the plots at $E_{e}$ = 30eV are not smooth for $m_{l}$ = 1 and 3 (see green and maroon curves). Unlike the odd OAM's, the angular distribution of TDCS remains unchanged of even OAM values. However, an increase in ejected energy leads to an enhanced forward peak and a suppressed backward peak as the sole effect (see Blue and indigo curves).
	From Table 1, we can see in detail the effects of the exchanged photons and different values of OAM. We observe that the ratio of TDCS from forward to backward direction varies with OAM. The ratio is higher for odd OAM values and increases with OAM (look at values for $m_l$ = 1 and $m_l$ = 3). It is seen to increase with the decrease in the value of $l$.

	\begin{center}
		\begin{table}[h]
			\label{table:tablenos}
			\begin{tabular}{ c | c |  c | c } 
				\hline
				\hline
				OAM \ & TDCS (0$^\circ$) / TDCS (180$^\circ$)  (l = 1)\ & TDCS (0$^\circ$) / TDCS (180$^\circ$) (l = 0) \ & TDCS (0$^\circ$) / TDCS (180$^\circ$) (l = -1) \\ [0.5ex] 
				\hline\hline
				$m_l$ = 1 \ & 6.0211 \ & 6.5674 \ & 6.6968 \\ 
				
				$m_l$ = 2 \ & 1.7894 \ & 1.8904 \ & 1.8411\\
				
				$m_l$ = 3 \ & 28.4957 \ & 34.0278 \ & 39.6002\\ 
				
				$m_l$ = 4 \ &  1.9604\ & 2.0857\ & 2.0409\\[1ex]
				\hline\hline 
			\end{tabular}
			\caption { First Born triple differential cross-section TDCS (a.u.) in the forward direction $0^\circ$ to the backward direction $180^\circ$ corresponding to the laser-assisted twisted electron impact ionization of hydrogen atom, for different values of OAM from 1 to 4 with the difference of unity. The laser frequency is 1.17eV, and the electric field strength is $10^{7} V/m$, and with the absorption of one photon, i.e. l = 1, no net transfer of photons (l = 0) and for emission of one photon ( l = -1 ). The incident energy $E_{i}$ = 500eV, the ejected energy $E_{e}$ = 30eV and the scattering angle $\theta_{s}$ = 100mr. The polarization direction of the electric field $\epsilon_{0}$ is parallel to the incident momentum $k_i$.}
		\end{table}
	\end{center}
	
	\begin{figure*}[htp!]
		\begin{tabular}{ccc} 
			\includegraphics[width=6.00cm]{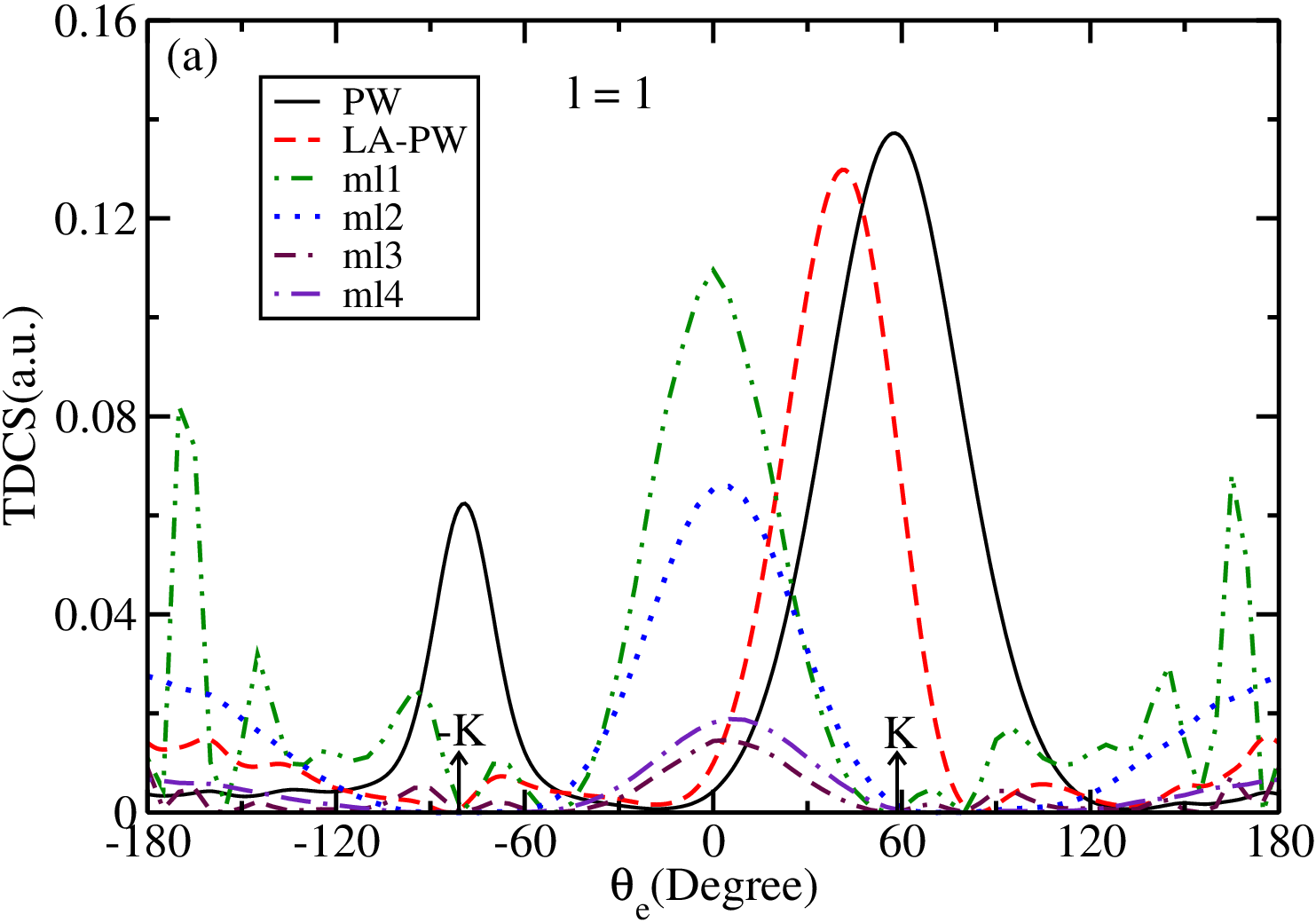}\ & \includegraphics[width=6.00cm]{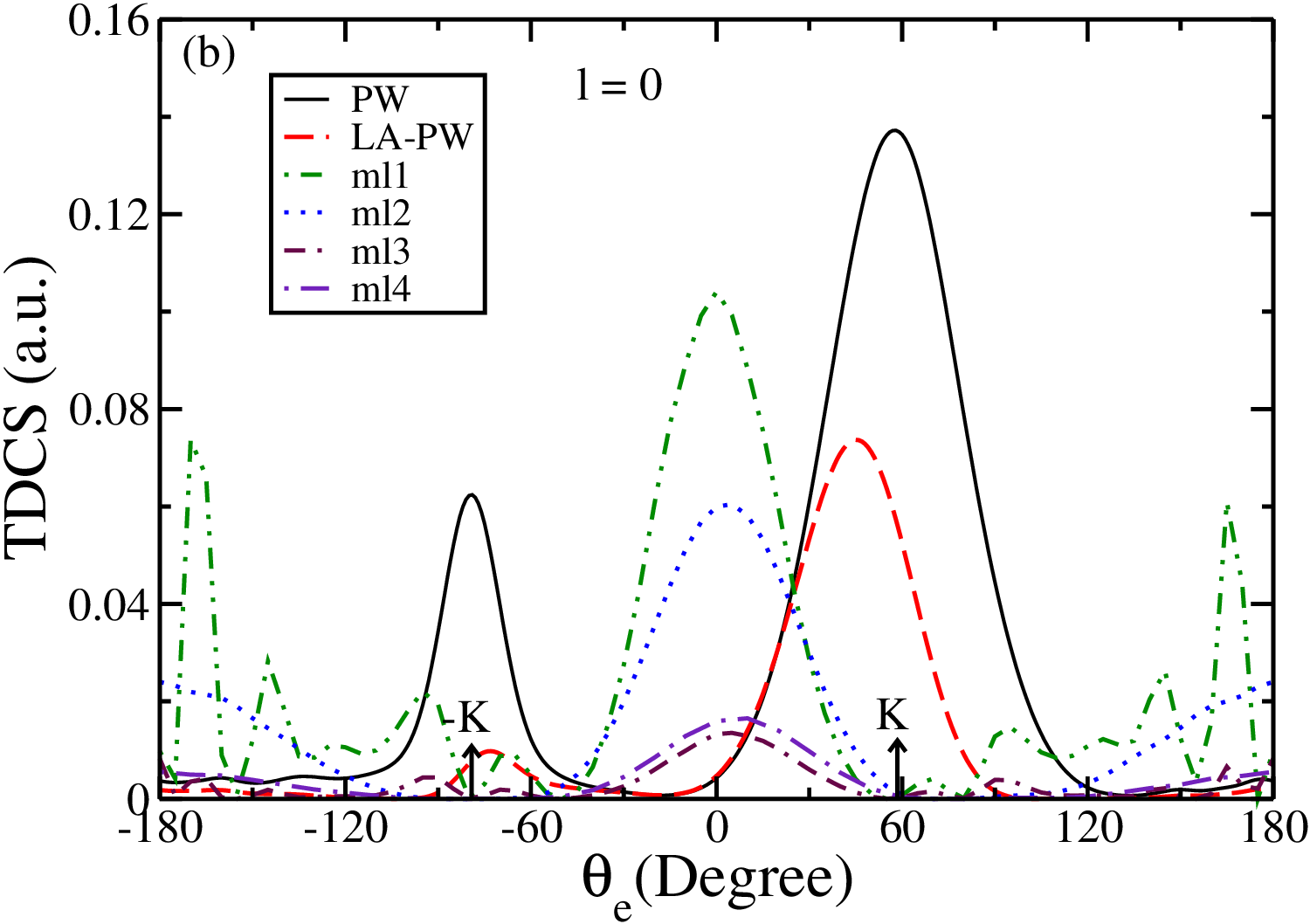}
			& \includegraphics[width=6.00cm]{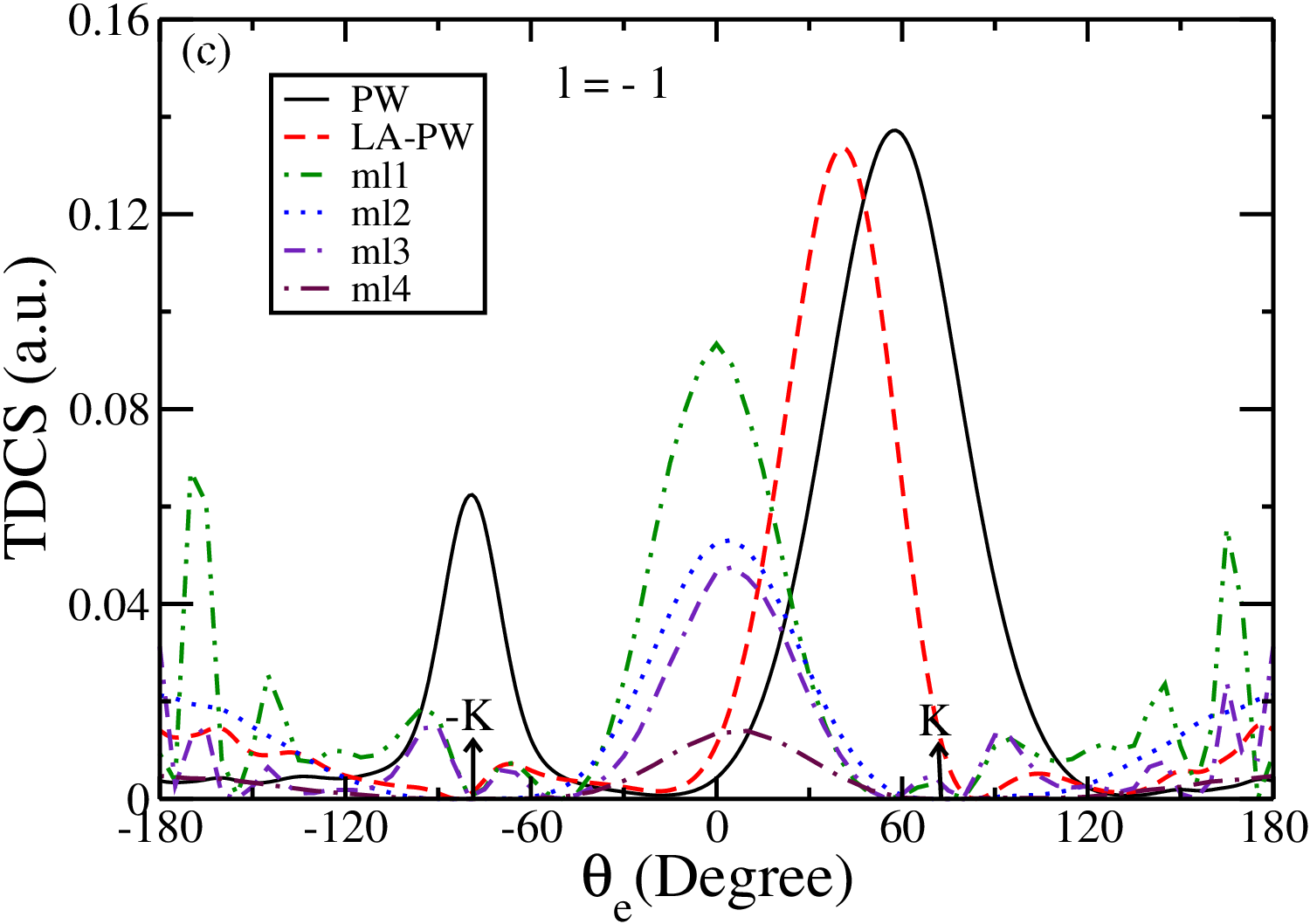}
		\end{tabular}
		\caption{Same as the figure 1 except the kinematics $E_e$ = 40eV. The magnitude of the TDCS for odd $m_l$ scaled up by 1000, for even $m_l$ by 100, and for LA-PW scaled up by a factor of 100, to compare with the PW results.}   
		\label{fig:3}
	\end{figure*}
	In figure \ref{fig:3} we plot TDCS for the ejeced electron energy $E_e$ = 40eV. For the PW calculations, we observe that the intensity of the binary peak decreases while that for the recoil peak increases (see fig 3 black solid curve). The magnitude of the TDCS for no net transfer of photons is relatively smaller in LA-PW (see fig 3(b) red dashed curve) compared to the absorption ($l$ = 1)(see fig 3(a) red dashed curve) and emission ($l$ = -1) (see fig 3(c) red dashed curve) of a single photon. For $l$ = $\pm1$ recoil peak diminishes, while a small recoil peak is observed for $l$ = 0 (see red dashed curve fig 3). Furthermore, in the case of LA-TEB, significant changes occur in the angular distribution of the TDCS for odd values of the OAM at ejected energies 40 eV and 50 eV. As depicted in Figure 3, for OAM $m_l$ = 1 (see green curve fig 3) and 3 (see the maroon curve), the peaks are no longer smooth but split into multiple smaller peaks with varying amplitudes.
	
	For even values of the OAM, there is no alteration in the angular distribution of the TDCS (see curves blue and indigo in fig 3). However, with increased ejected electron energy, the forward peaks enhanced and the backward peak suppressed significantly, i.e.
	the ratio of the TDCS from the forward direction to the backward direction increases as the ejected energy increases, as evident from Tables 2 and 3. 
	\begin{center}
		\begin{table}[h]
			\begin{tabular}{ c | c |  c | c } 
				\hline
				\hline
				OAM \ & TDCS (0$^\circ$) / TDCS (180$^\circ$)  (l = 1)\ & TDCS (0$^\circ$) / TDCS (180$^\circ$) (l = 0) \ & TDCS (0$^\circ$) / TDCS (180$^\circ$) (l = -1) \\ [0.5ex] 
				\hline\hline
				$m_l$ = 1 \ & 10.0009 \ & 10.6558 \ & 10.0684 \\ 
				
				$m_l$ = 2 \ & 2.3740 \ & 2.4983 \ & 2.4403 \\
				
				$m_l$ = 3 \ & 1.6078 \ & 1.6161 \ & 1.4897 \\ 
				
				$m_l$ = 4 \ & 2.7329  \ & 2.8859  \ & 2.8696  \\[1ex]
				\hline\hline 
			\end{tabular}
			\caption { Same as table 1, but the ejected energy is taken to be 40 eV.}
		\end{table}
	\end{center}
	\begin{figure*}[htp!]
		\begin{tabular}{ccc}
			\includegraphics[width=6.00cm]{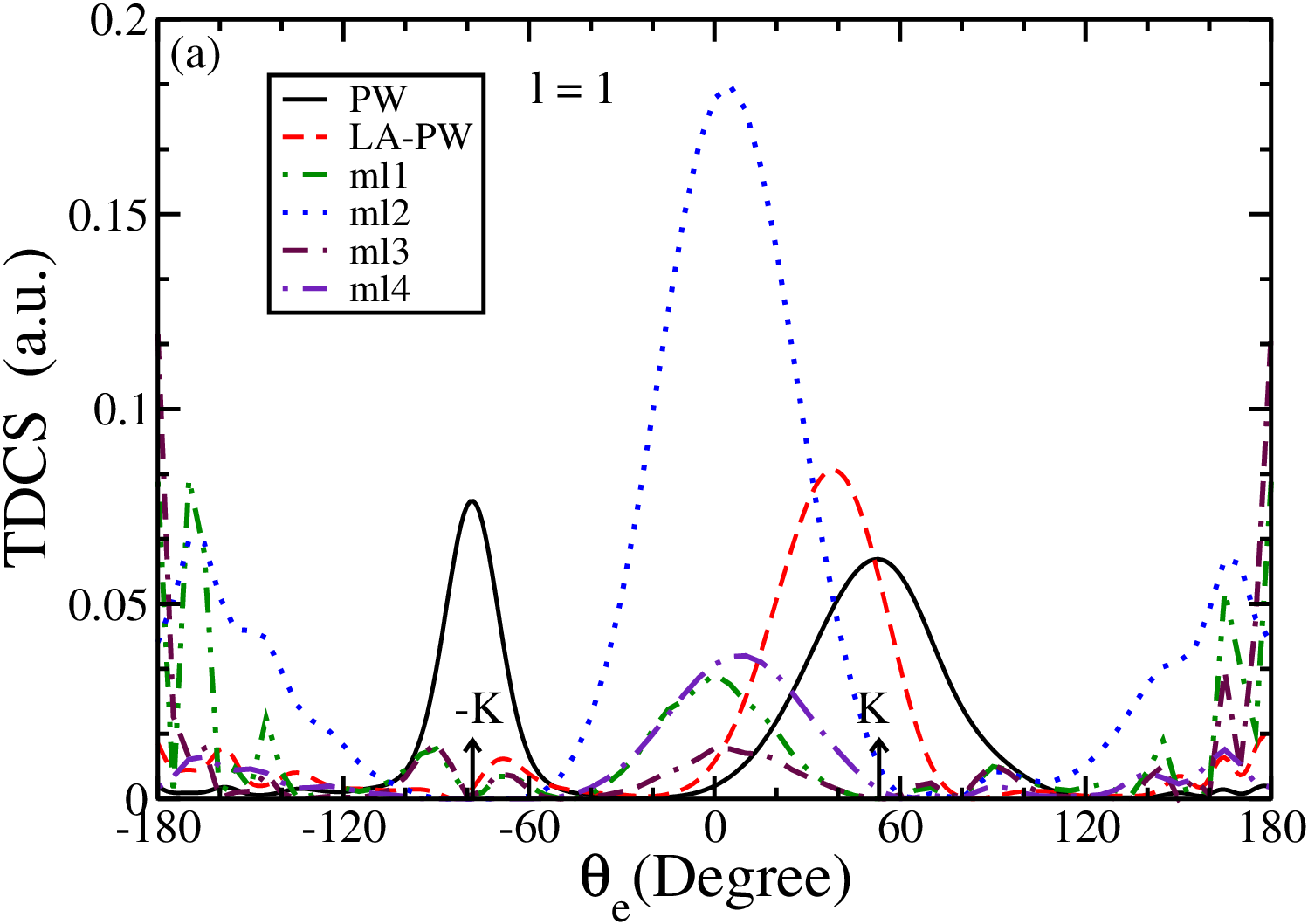}\ & \includegraphics[width=6.00cm]{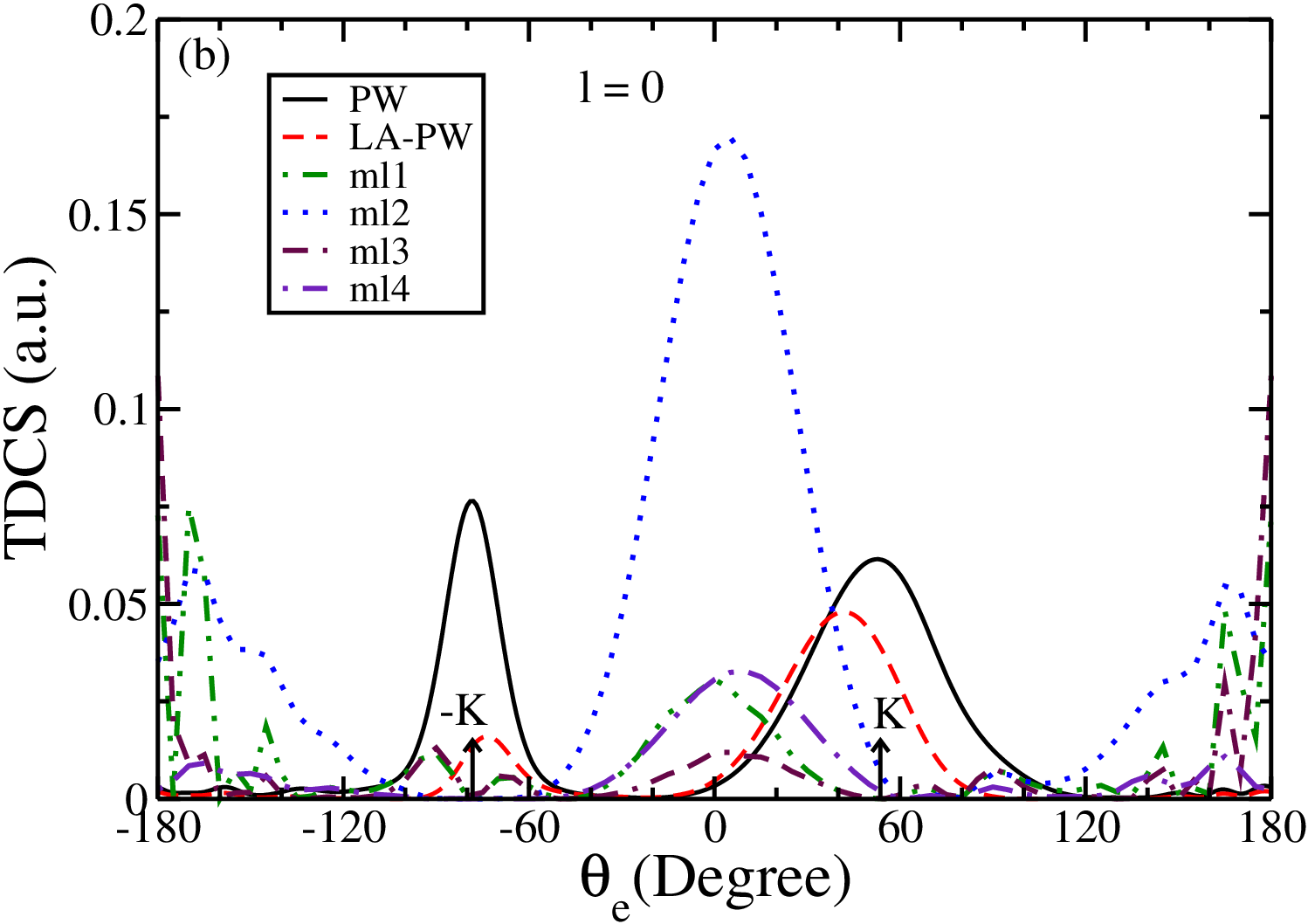}
			& \includegraphics[width=6.00cm]{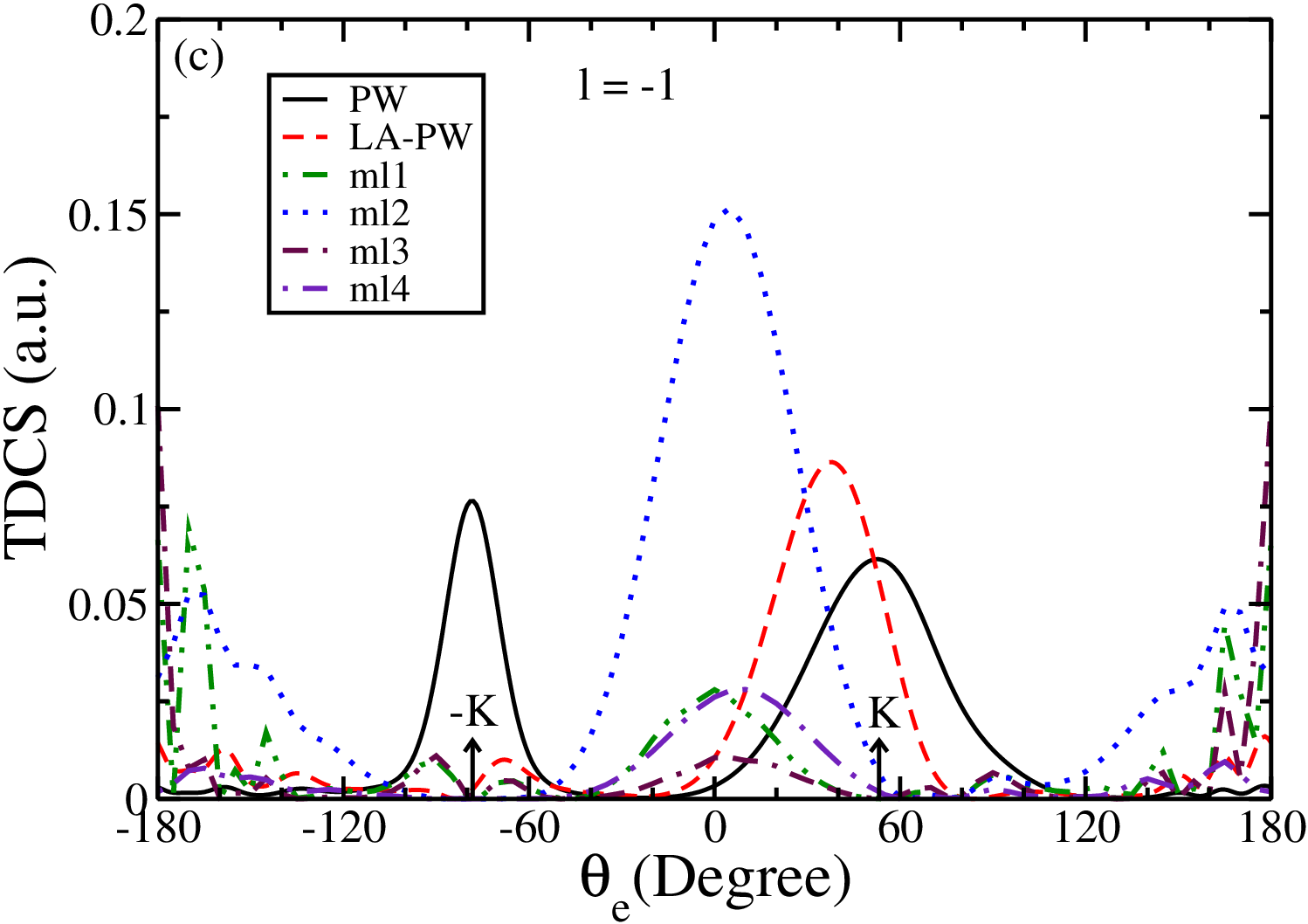}
		\end{tabular}
		\caption{Same as the figure 1 except the kinematics $E_e$ = 50eV. The TDCS is plotted for different values of $m_l$ and $l$. The magnitude of the TDCS for different values of OAM  is scaled up by a factor of 1000 to compare with the PW results.}   
		\label{fig:4}
	\end{figure*}
	In Figure \ref{fig:4}, the TDCS is plotted for the ejected energy 50eV. For plane wave, the magnitude of the binary peak decreases, and the peak shifts counterclockwise (see Black solid curve) . Futhermore, the magnitude of TDCS for the recoil peak increases and becomes dominant over the binary peak (see black solid curve).
On contrary of this, the LA plane wave angular profile of the TDCS remains largely unchanged with the increased ejected electron energy (40eV to 50eV), except for the reduction in the absolute magnitude of TDCS (see red dashed curve).
	At higher ejected energy 50eV, the angular distribution of LA-TEB changes drastically. 
	We observed a very shallow peak in the forward direction for odd OAM numbers $ m_l$'s 1 and 3, along with multiple small peaks of varying magnitudes in different directions (see green and maroon curves in fig 4). For $m_l$ = 2 and 4 also, a small distortion is noticed at this ejected electron energy (see Blue and Indigo curves). Also with the increasing ejected energy; the TDCS  magnitude difference increases between $m_l$ = 2 and 4 compared $m_l$ = 1 and 3.\\
	Upon evaluating the impact of transferred photons from the laser field to the electron-atom system, we noted a decrease in the magnitude of TDCS with a reduction in the number of transferred photons ( $l$ = 0, $\pm1$) while angular distribution of TDCS remains same. This trend remains consistent across Figures 1 to 4, encompassing various ejected energies used.

	\begin{center}
		\begin{table}[h]
			\begin{tabular}{ c | c |  c | c } 
				\hline
				\hline
				OAM \ & TDCS (0$^\circ$) / TDCS (180$^\circ$)  (l = 1)\ & TDCS (0$^\circ$) / TDCS (180$^\circ$) (l = 0) \ & TDCS (0$^\circ$) / TDCS (180$^\circ$) (l = -1) \\ [0.5ex] 
				\hline\hline
				$m_l$ = 1 \ & 0.3936 \ & 0.4216 \ & 0.4210 \\ 
				
				$m_l$ = 2 \ & 4.4702 \ & 4.7227 \ & 4.7606 \\
				
				$m_l$ = 3 \ & 0.1087 \ & 0.1099 \ & 0.1048 \\ 
				
				$m_l$ = 4 \ & 11.7611  \ & 13.2900 \ & 14.5384 \\[1ex]
				\hline\hline 
			\end{tabular}
			\caption { Same as table 1, but the ejected energy is taken to be 50 eV.}
		\end{table}
	\end{center}
	\begin{figure*}[htp!]
		\begin{tabular}{ccc}
			\includegraphics[width=6.00cm]{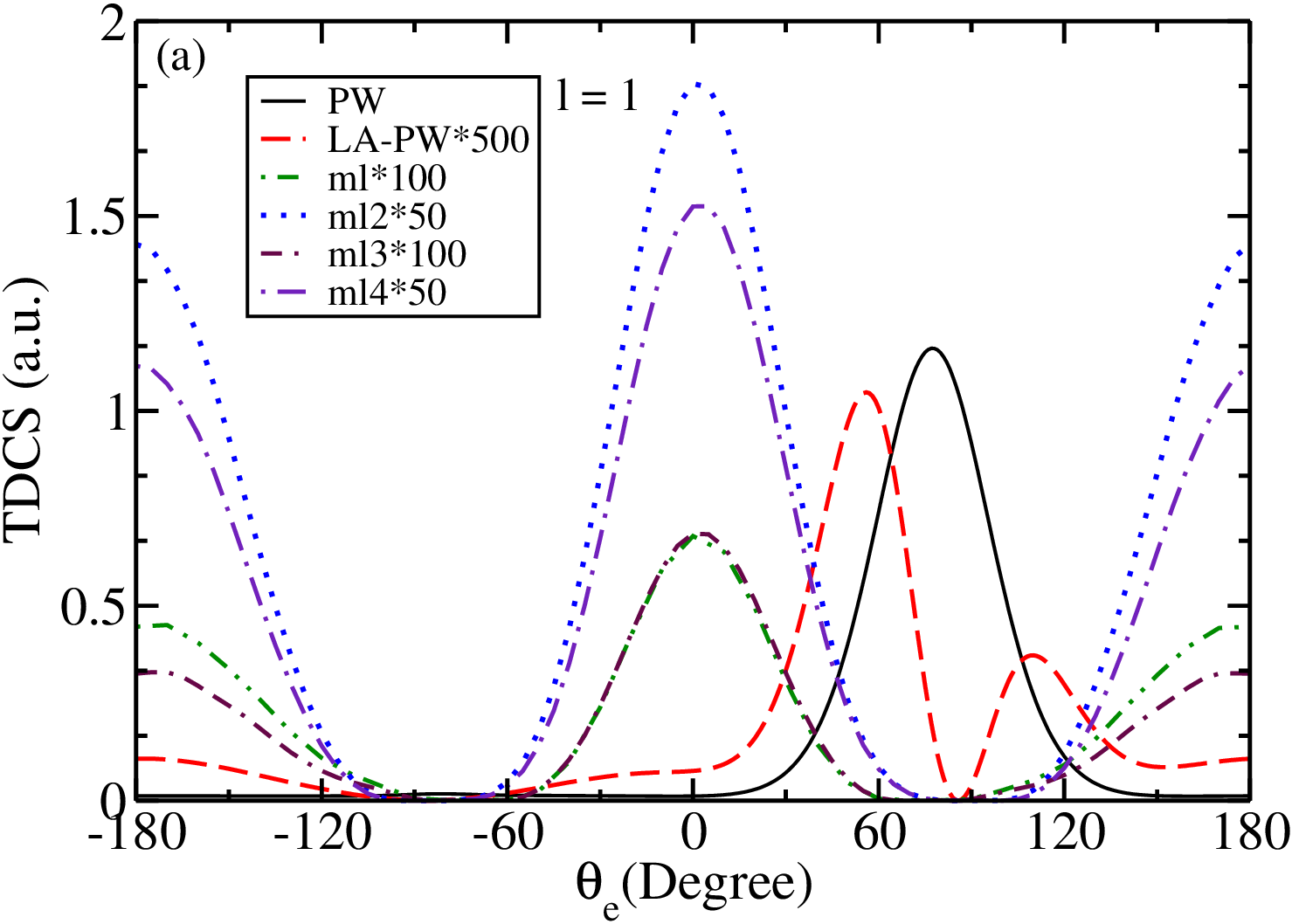}\ & \includegraphics[width=6.00cm]{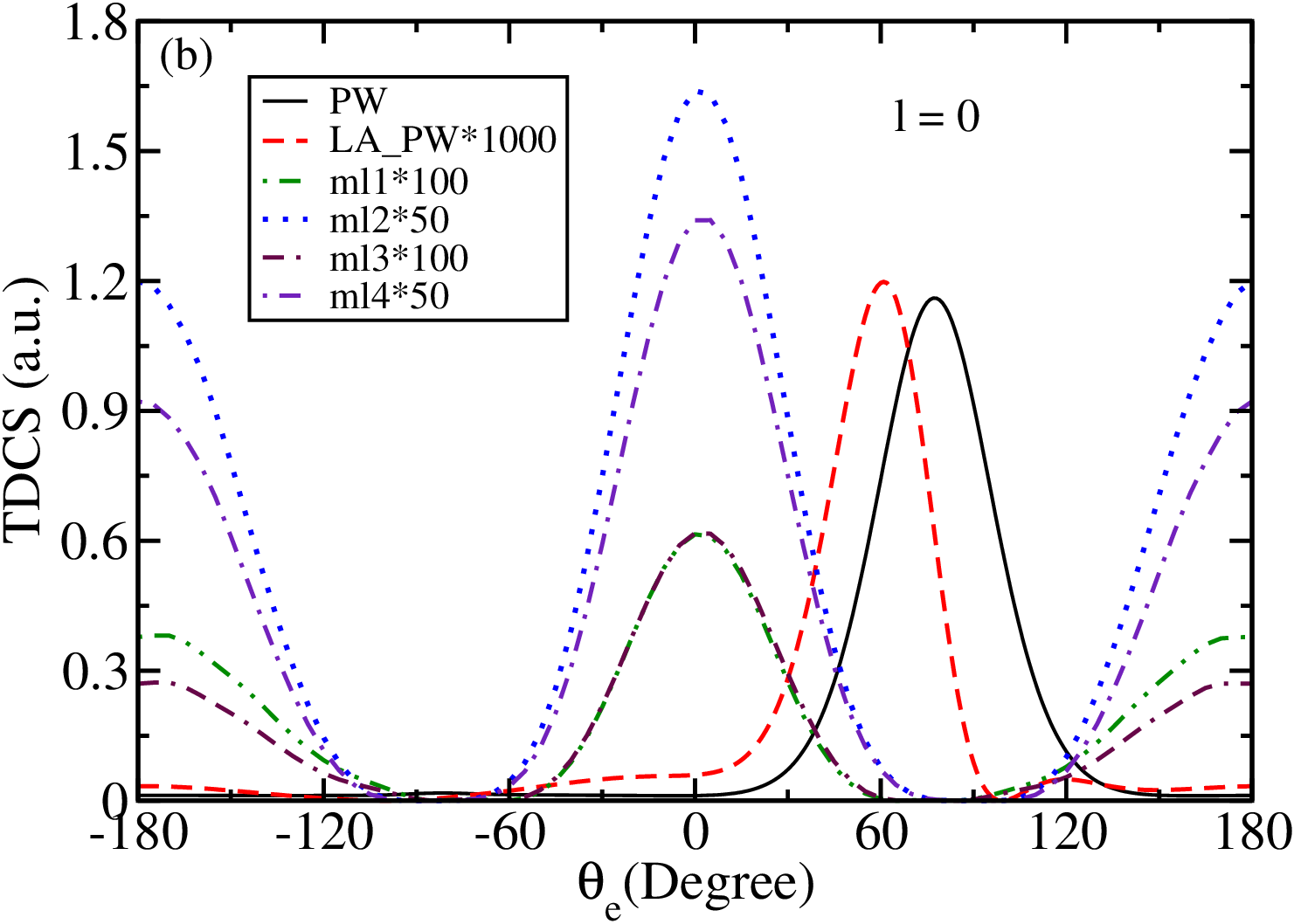}
			& \includegraphics[width=6.00cm]{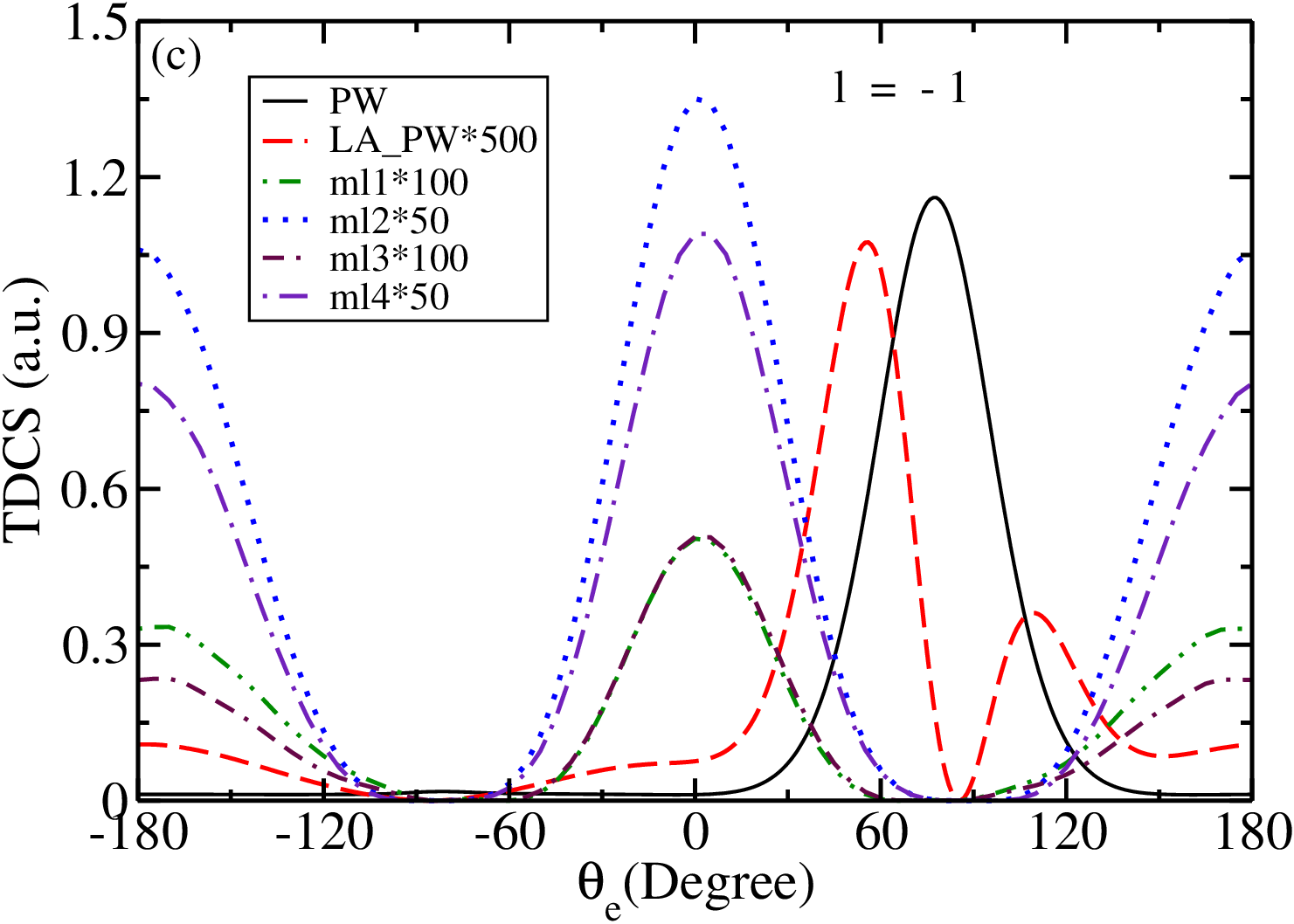}
		\end{tabular}
		\caption{Same as the figure 1 except the kinematics $E_i$ = 1000eV and $E_e$ = 20eV, TDCS is plotted for different valus of $m_l$ and $l$. The different scaling factors are shown in the each fram for better comparison with PW results mentioned in the panel.}   
		\label{fig:5}
	\end{figure*}
	We also analyze the effect of higher incident energy $E_i$ on the angular profile of TDCS for plane wave, LA plane wave, and LA-TEB considering different values of OAM ($m_l$ = 1, 2, 3, and 4) in the presence of a laser source for different ejected energies. We plot the angular profiles of various TDCS in the figures \ref{fig:5} - \ref{fig:8} for the different ejected electron energies and exchanged photons ($l$ = 0, $\pm$1).
	The TDCS's absolute magnitude increases with increasing incident energy for all three cases (PW, LA-PW, and LA-TEB). In our study of \textbf{PW}, we noticed two peaks, a binary, and a recoil peak; with the increase of ejected electron energy, the recoil peak became more prominent (refers to figures \ref{fig:5} - \ref{fig:8} black solid curves). However, we observed only a binary peak for LA plane wave incidence, which exhibited a noticeable shift toward the incidence direction compared to the plane wave case. The angular profile of TDCS for the LA plane wave differs with the transfer of photons and no net transfer of photons. We also observed a recoil peak at higher ejected energy ($E_e$ = 40eV and 50eV)  (see red dashed curve in figure \ref{fig:7}(b) - \ref{fig:8}(b)) for l = 0. For the LA-TEB, as the incidence energy increases, the absolute magnitude of the TDCS also increases while the peak positions remain unchanged. As previously mentioned, the angular distribution of the TDCS differs significantly for even and odd values of the OAM. The magnitude of the TDCS is greater for even OAM values compared to odd values, and this discrepancy becomes more pronounced as the incident energy increases ( see blue, indigo, green and maroon curves in figures \ref{fig:1} and \ref{fig:5}). 
	 In figure \ref{fig:6}, with the increment in the ejected electron energy for odd OAM numbers 1 and 3 (see green and maroon curves in fig 6), we observed a number of small peaks along with the forward peak; conversely, for even values 2 (see curve blue) and 4 (see curve indigo), we observed smooth forward and backward peaks. For odd $m_l$ values 1 and 3, only the forward peaks are clearly observable in the resulting output when analyzing the effect of the highest ejected energy employed in this communication ($E_e$ 40 and 50eV). However, they appear slightly distorted for odd $m_l$ (refer to green and maroon curves in Figure's \ref{fig:7} and \ref{fig:8}). \\
	Finally we would like to mention that the trend followed by all the cases discussed above is, in the direction of plane-wave binary peak and recoil peak , we noticed minima for LA-TEB incidence.
	
	\begin{figure*}[htp!]
		\begin{tabular}{ccc}
			\includegraphics[width=6.00cm]{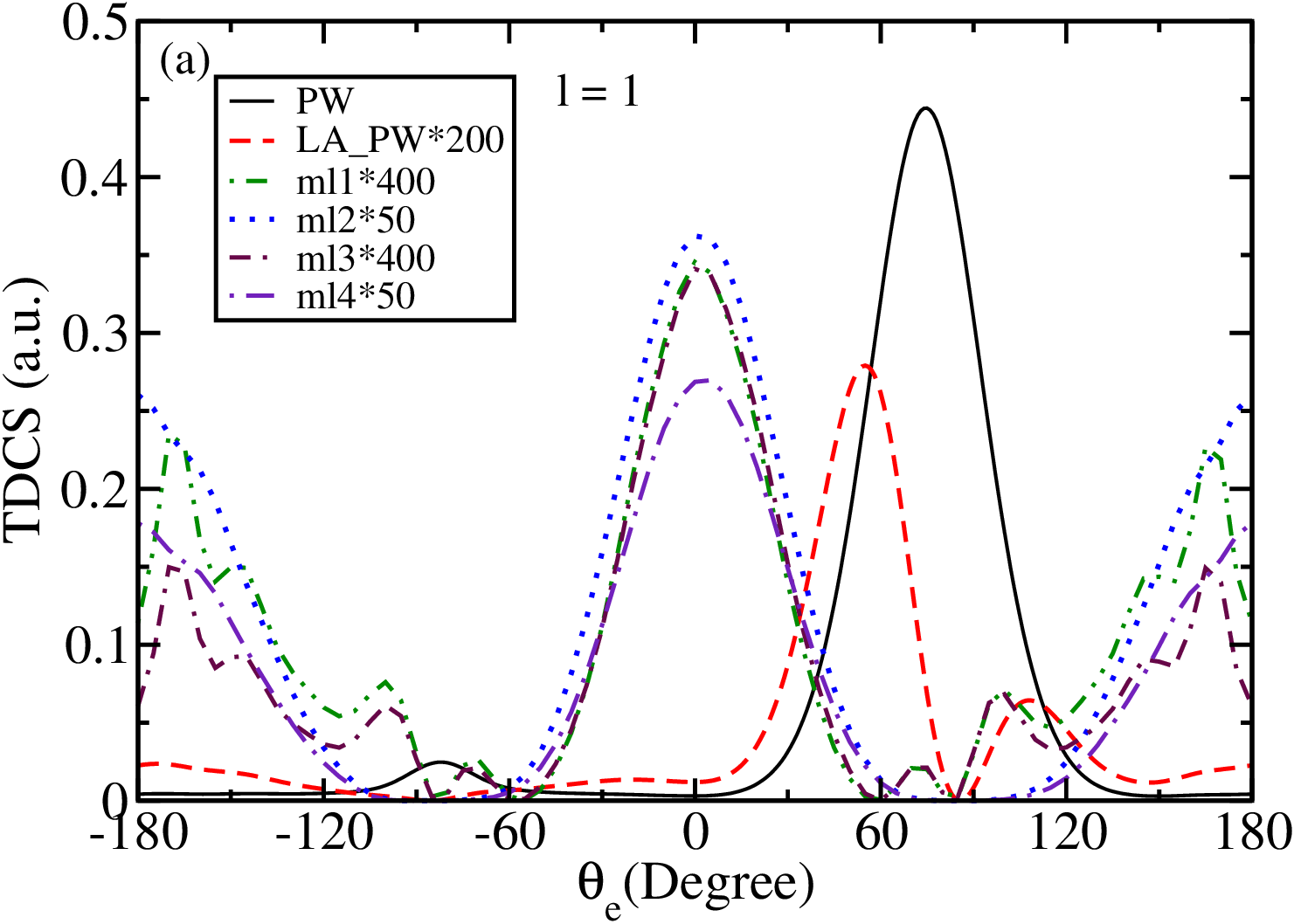}\ & \includegraphics[width=6.00cm]{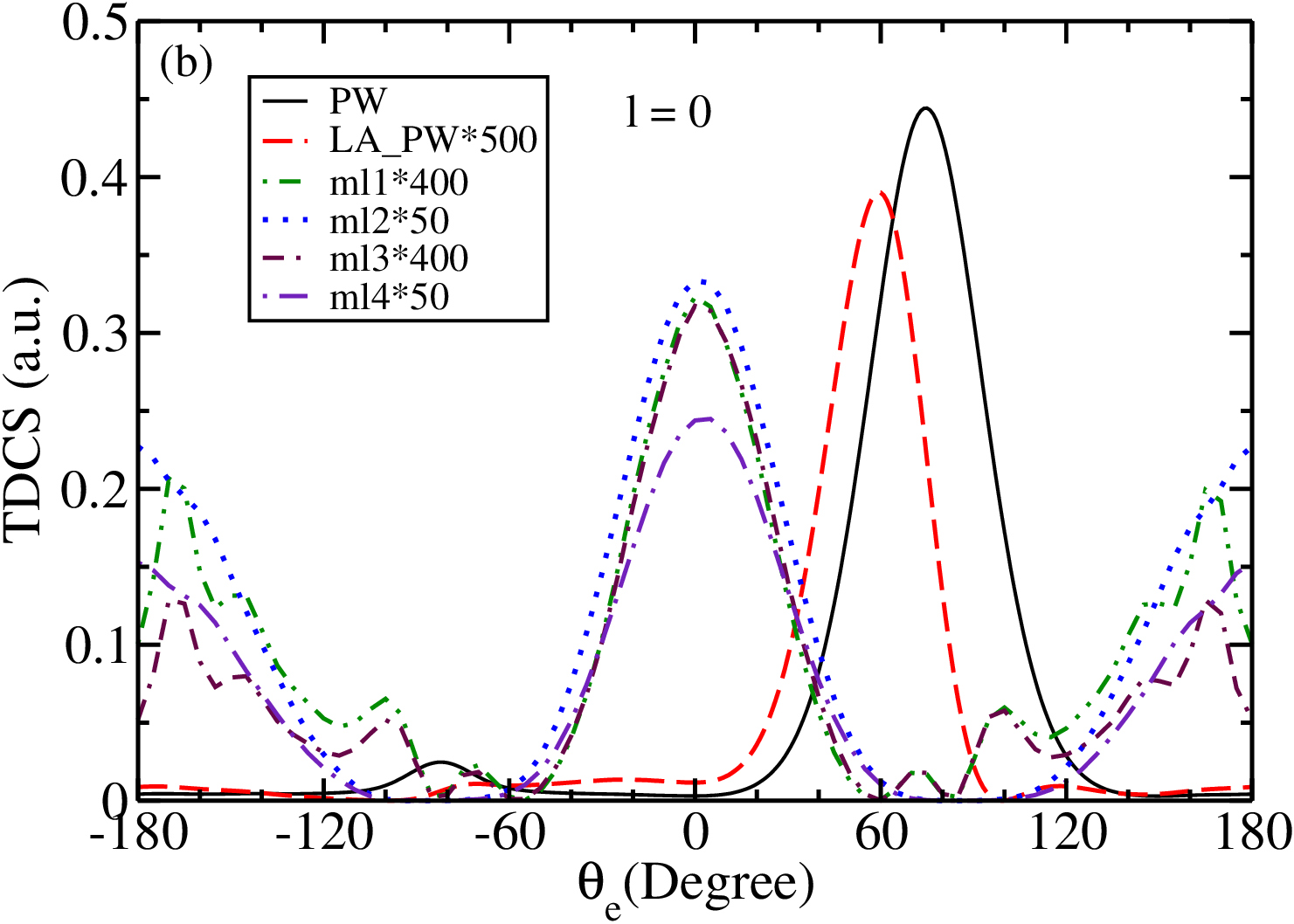}
			& \includegraphics[width=6.00cm]{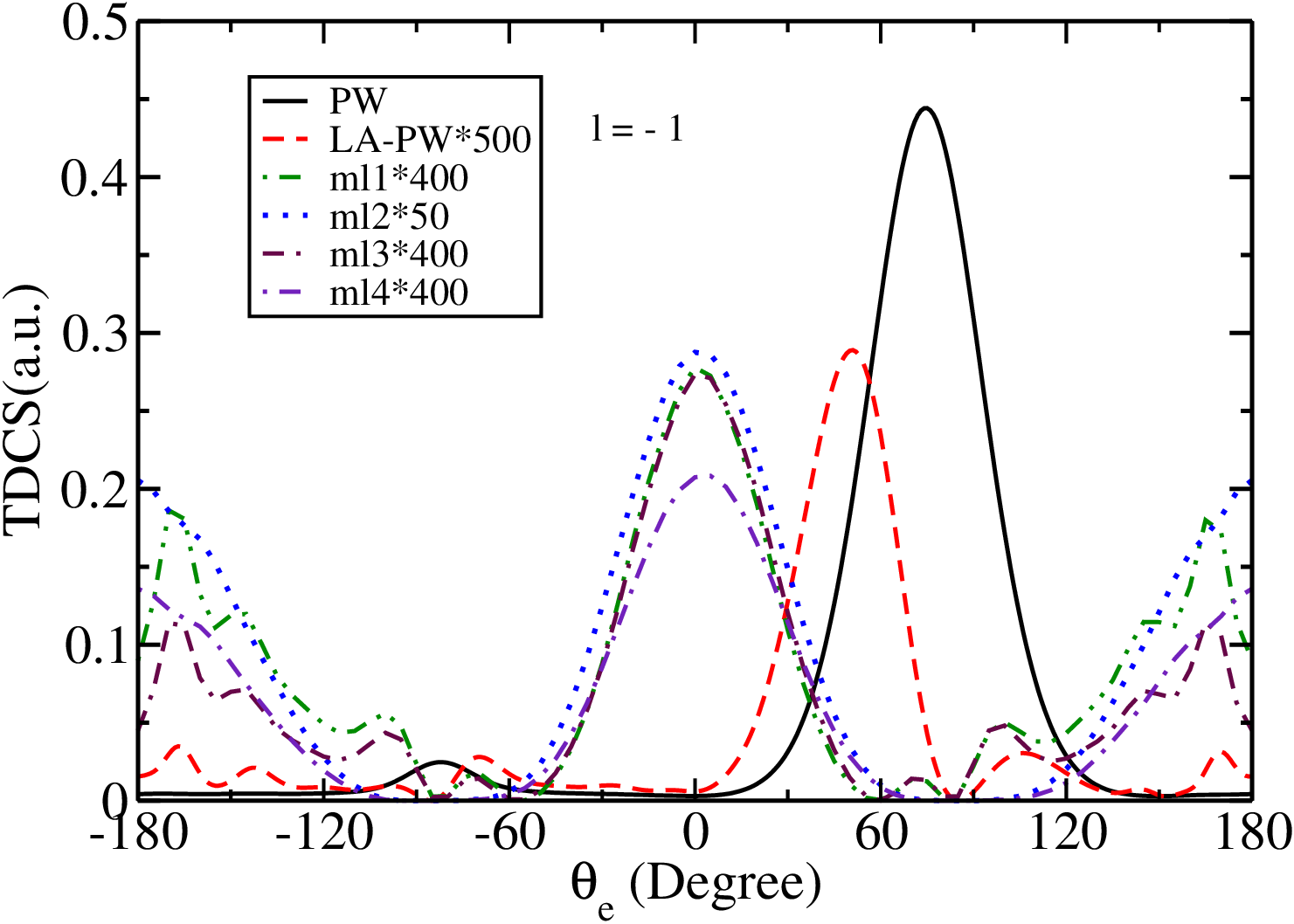}
		\end{tabular}
		\caption{Same as the figure 5 except the kinematics $E_e$ = 30eV. The different scaling factors are used for better comparison with PW results mentioned in the panel.}   
		\label{fig:6}
	\end{figure*}
				
				
				
	\begin{figure*}[htp!]
		\begin{tabular}{ccc}
			\includegraphics[width=6.00cm]{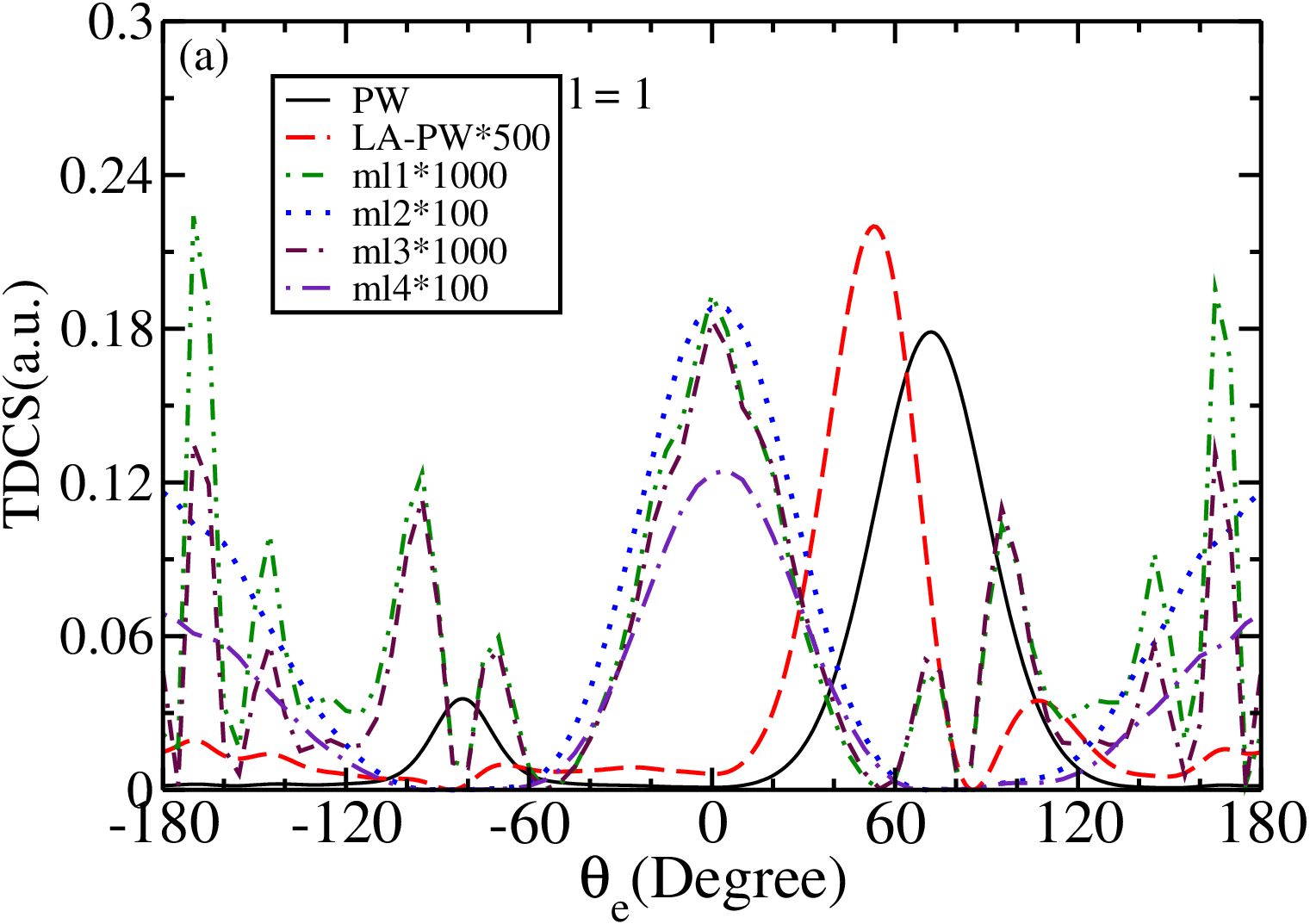}\ & \includegraphics[width=6.00cm]{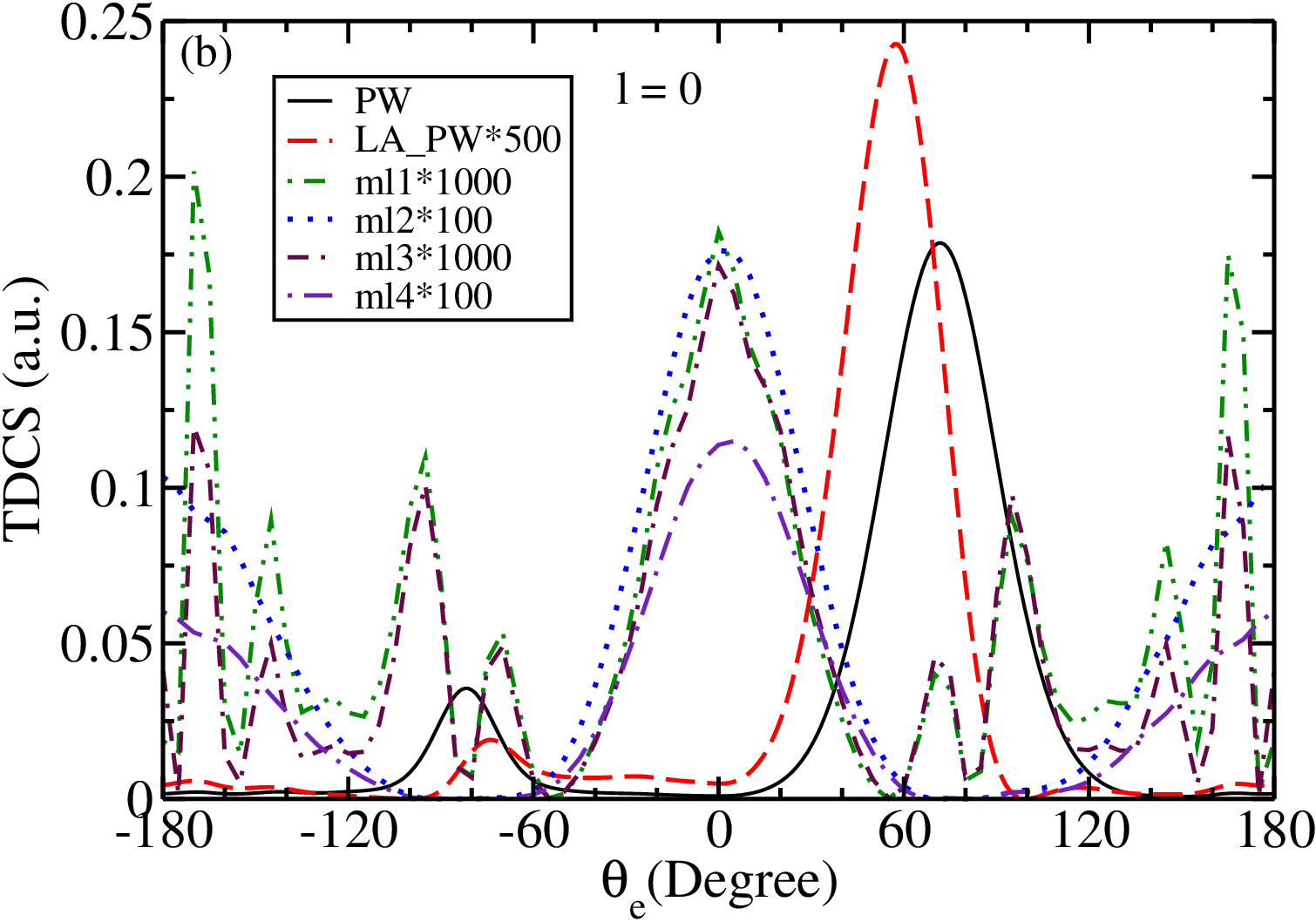}
			& \includegraphics[width=6.00cm]{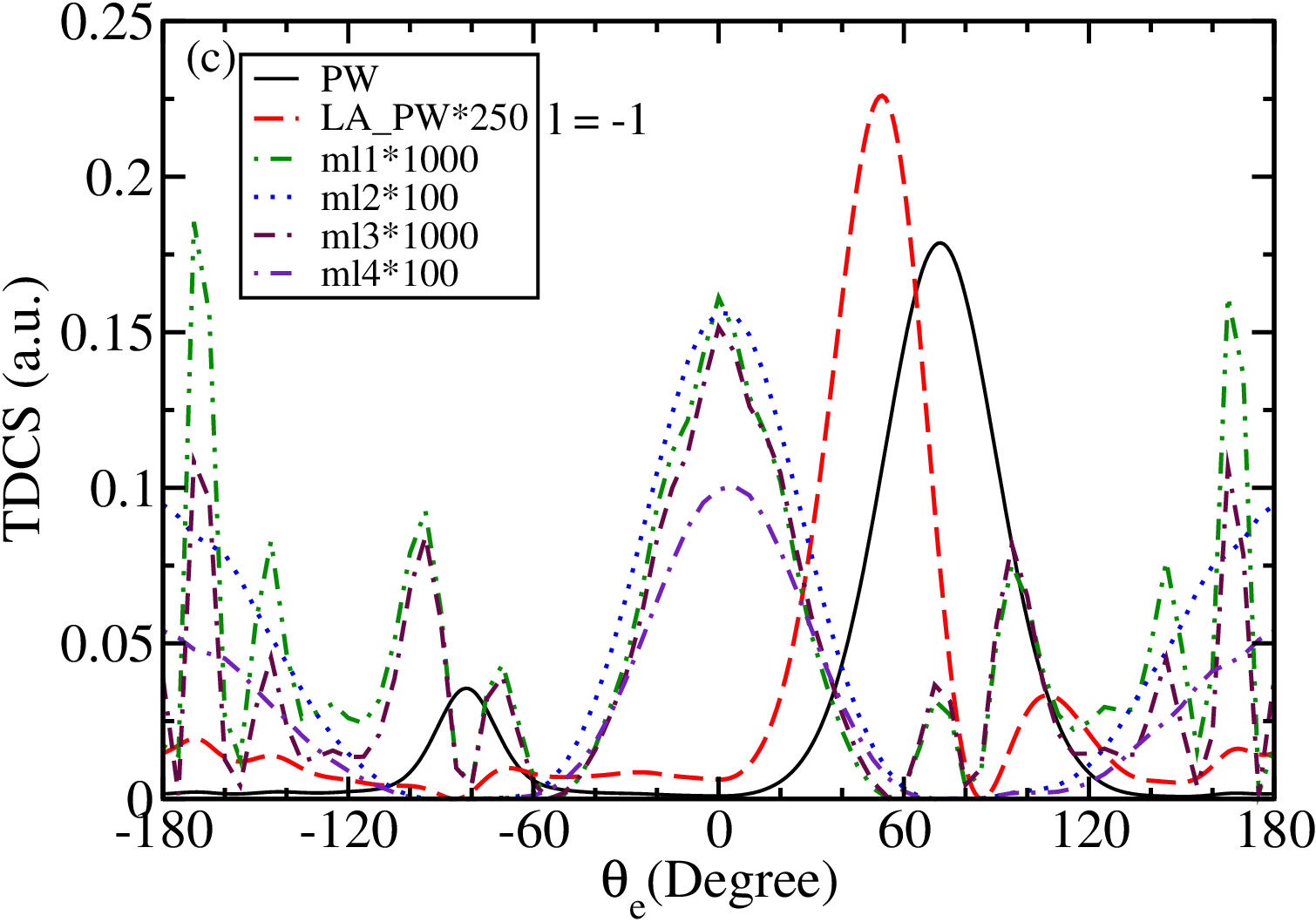}
		\end{tabular}
		\caption{Same as the figure 5 except the kinematics  $E_e$ = 40eV. The different scaling factors are used for better comparison with PW results mentioned in the panel. }
		\label{fig:7}
	\end{figure*}
				
				
				
	\begin{figure*}[htp!]
		\begin{tabular}{ccc}
			\includegraphics[width=6.00cm]{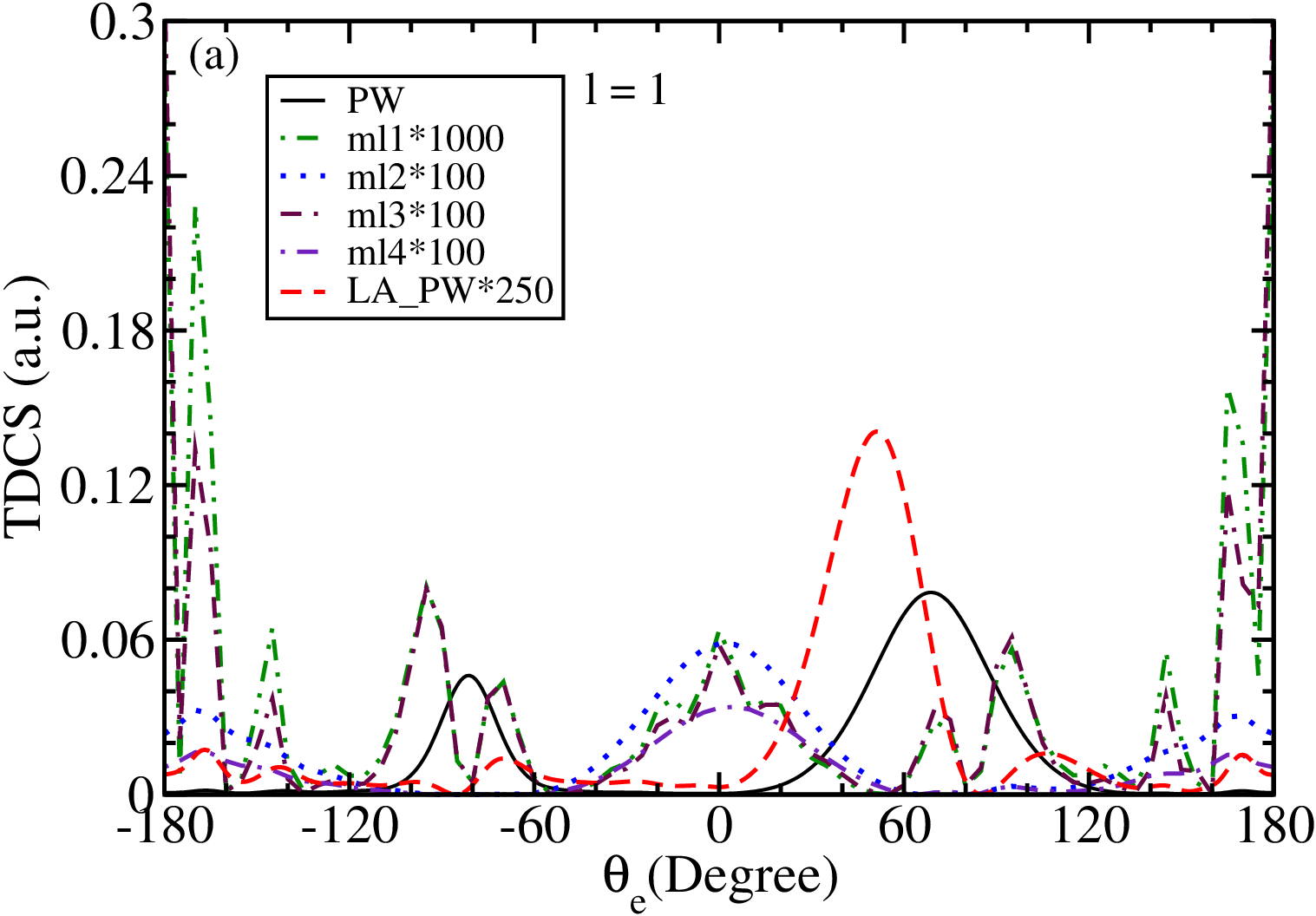}\ & \includegraphics[width=6.00cm]{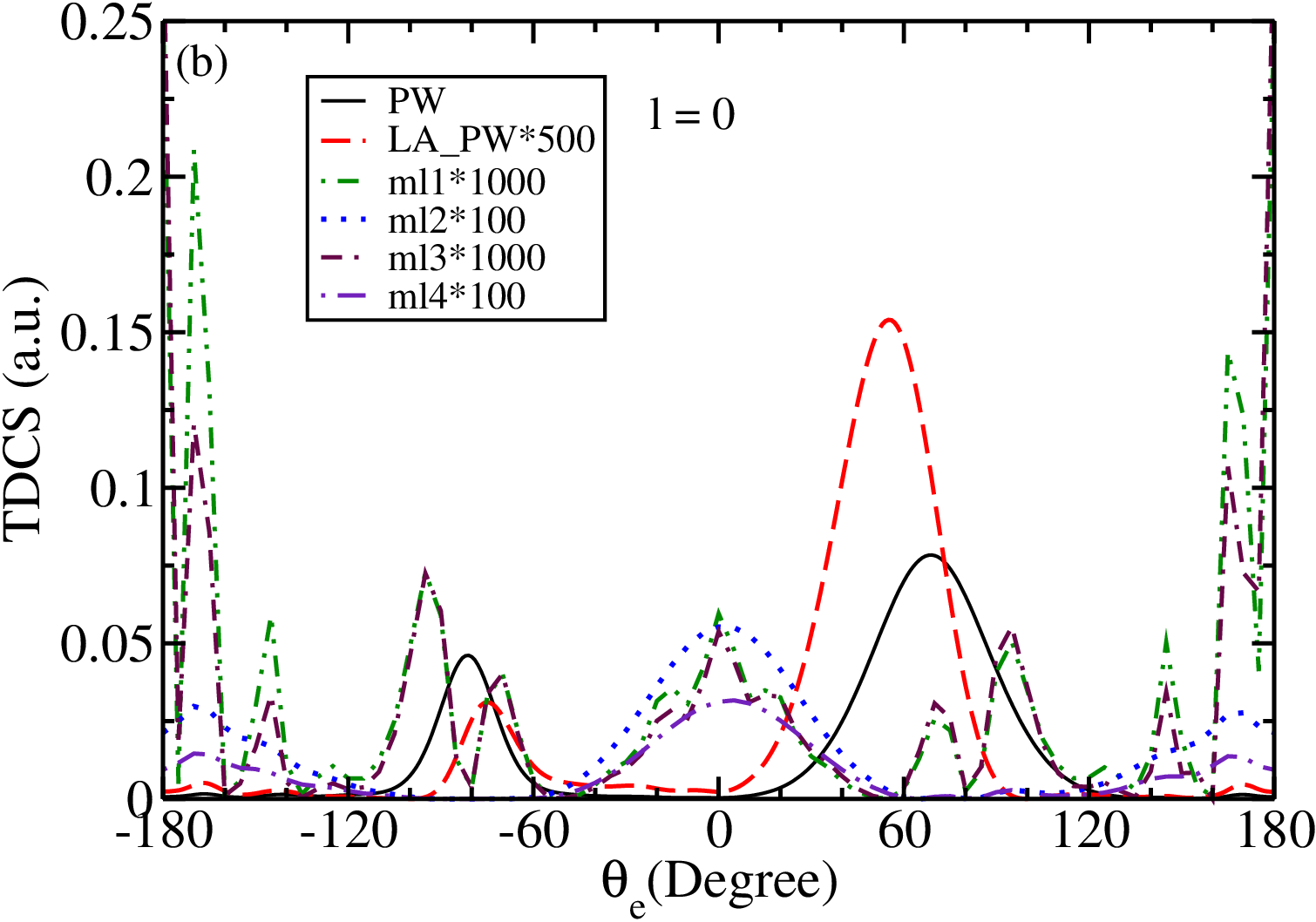}
			& \includegraphics[width=6.00cm]{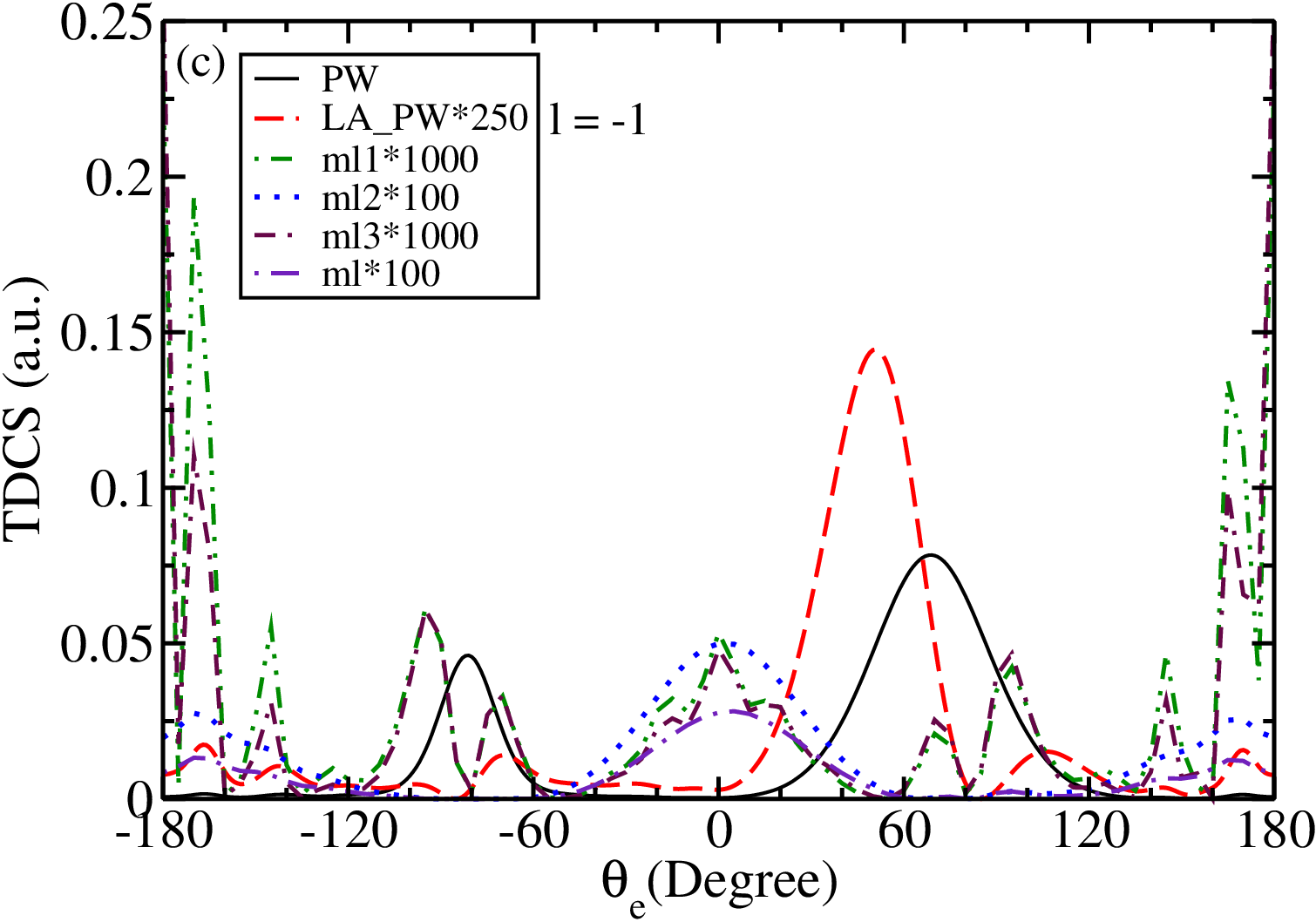}
		\end{tabular}
		\caption{Same as the figure 5 except the kinematics $E_e$ = 50eV. The different scaling factors are used for better comparison with PW results mentioned in the panel.}   
		\label{fig:8}
	\end{figure*}
				
				
				
			\subsection {LA TEB verses TEB without laser-filed}
\begin{figure*}[htp!]
				\begin{tabular}{cc}
					\includegraphics[width=8.00cm]{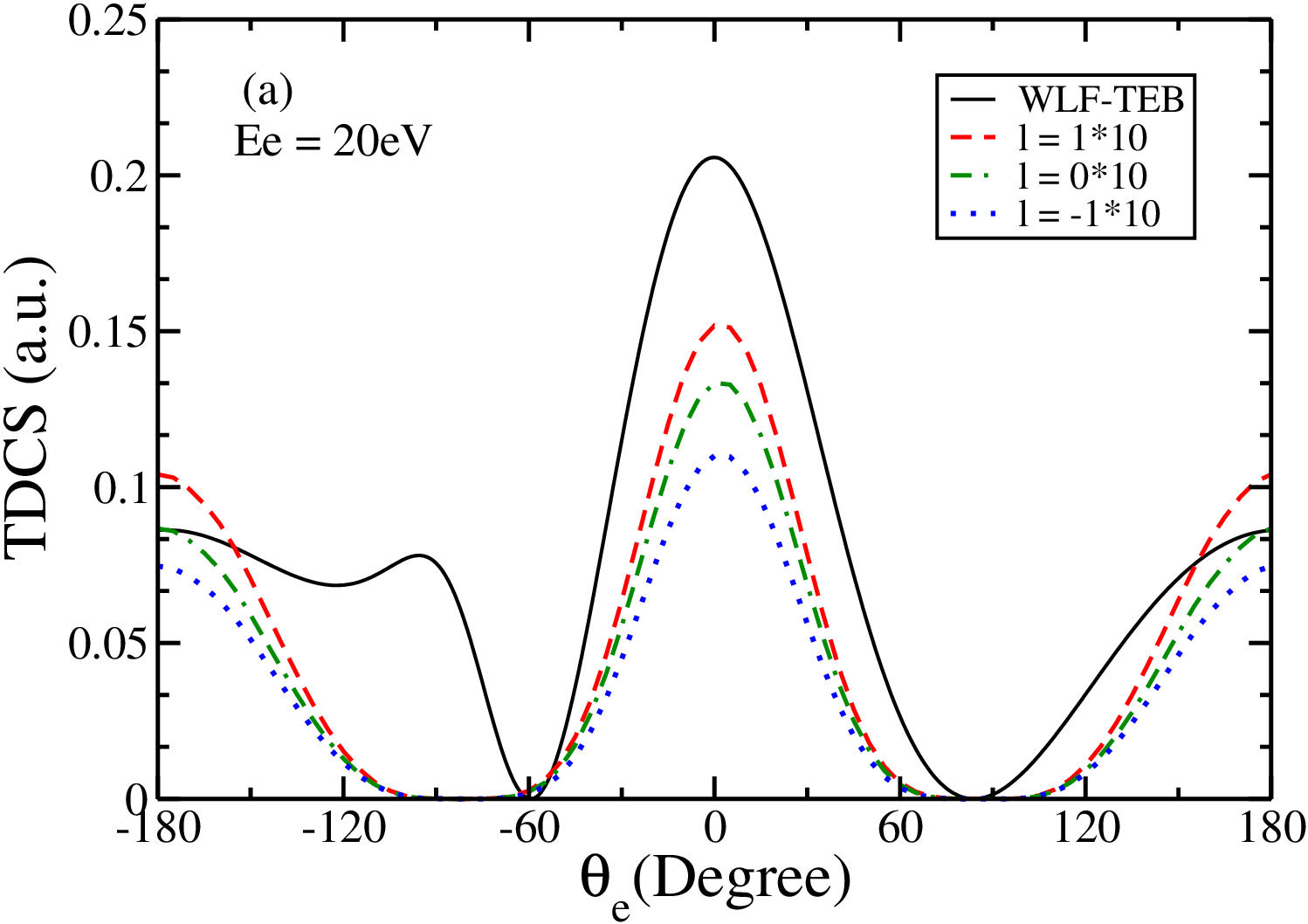}\ & \includegraphics[width=8.00cm]{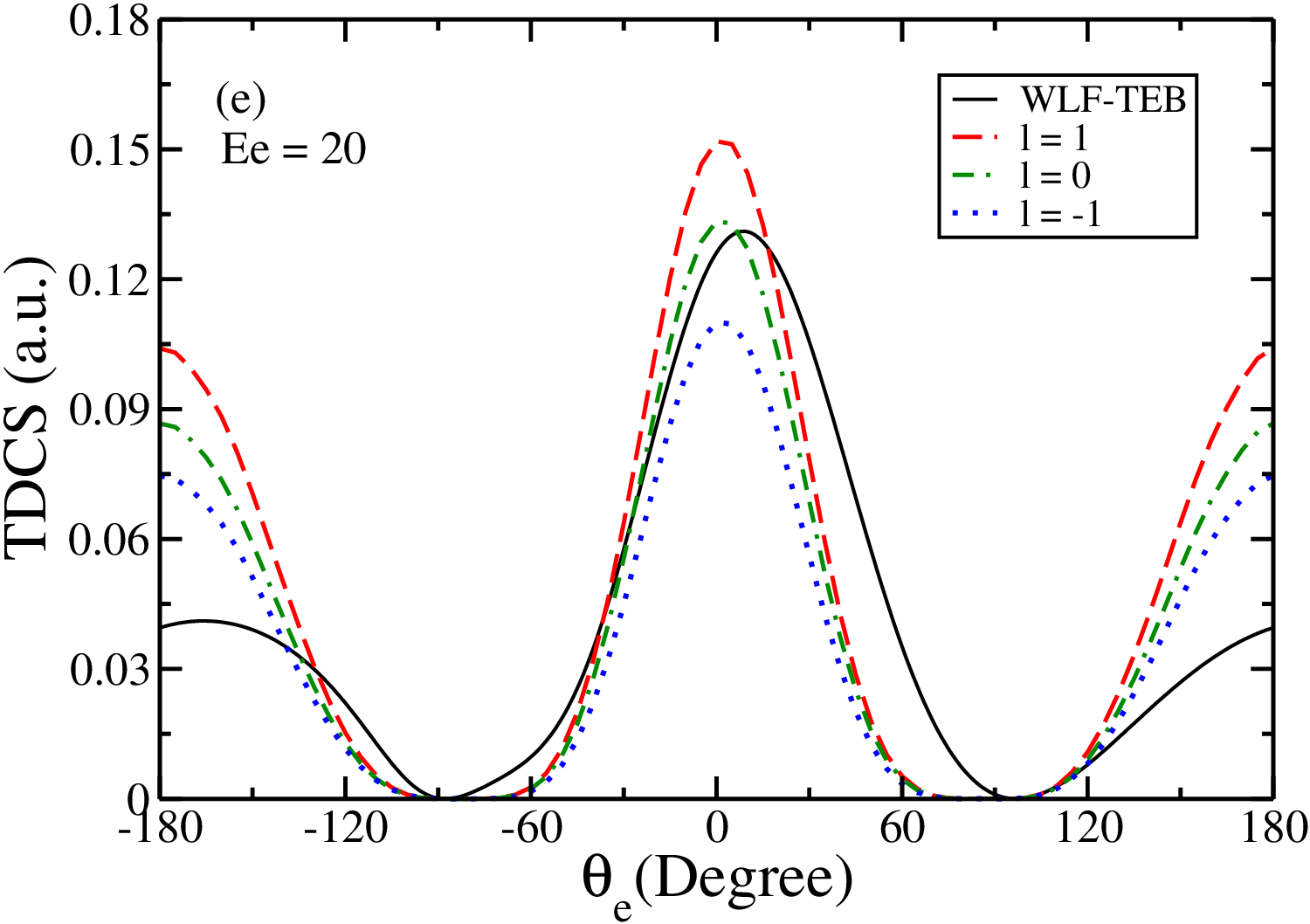}
				\end{tabular}

				\begin{tabular}{cc}
					\includegraphics[width=8.00cm]{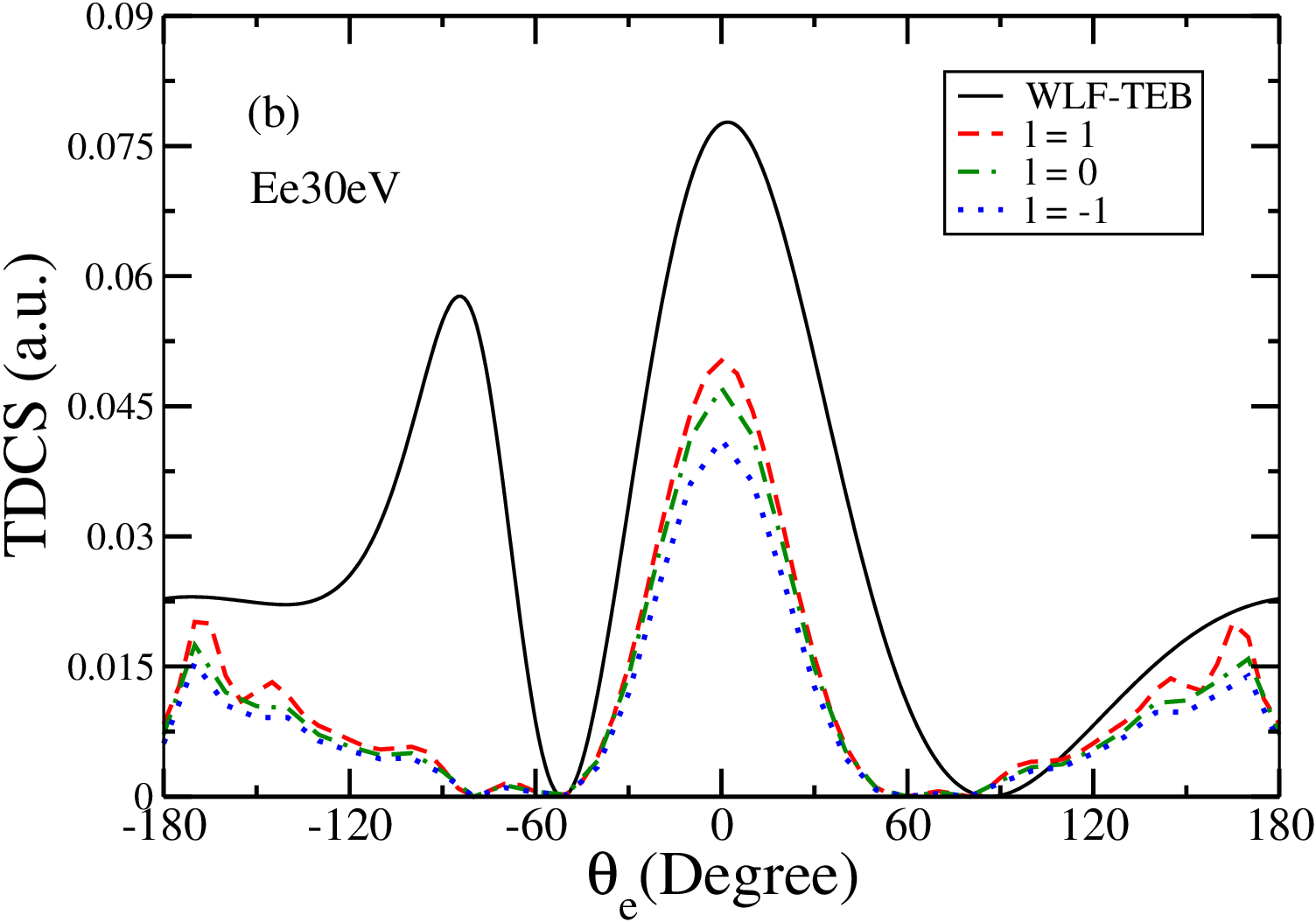}\ & \includegraphics[width=8.00cm]{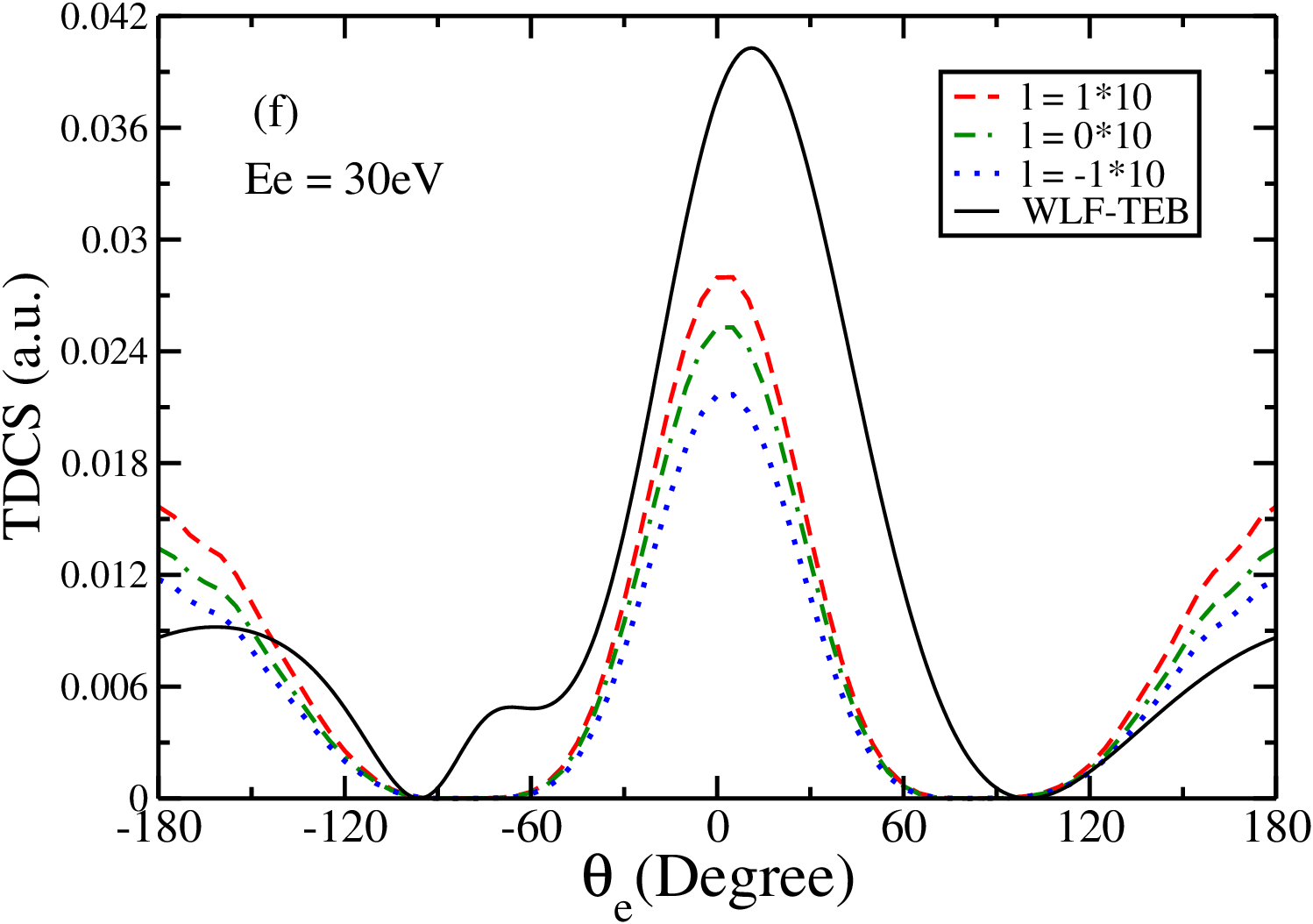}
					
				\end{tabular}
				
				\begin{tabular}{cc}
					\includegraphics[width=8.00cm]{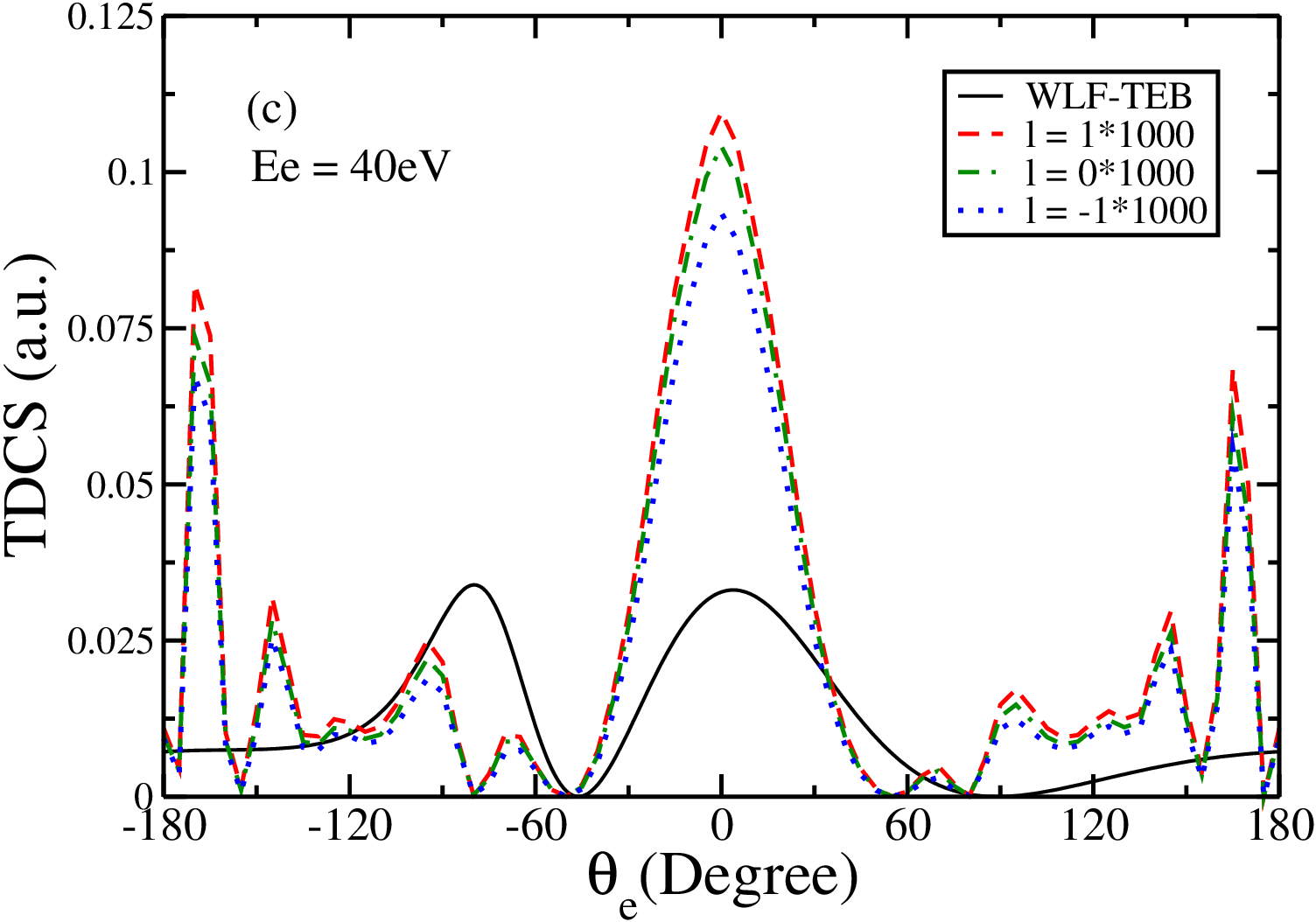}\ & \includegraphics[width=8.00cm]{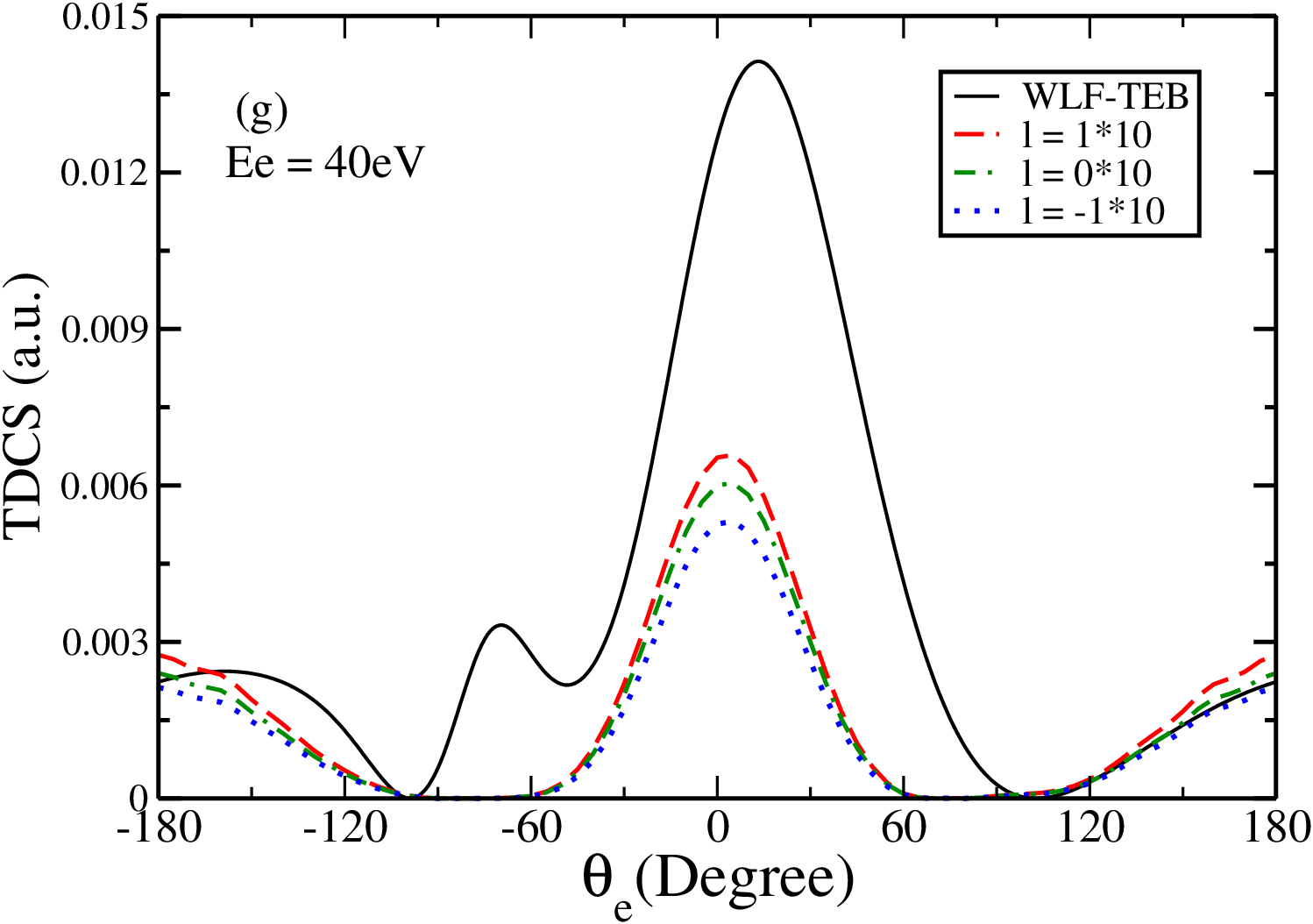}
					
				\end{tabular}
				
				\begin{tabular}{cc}
					\includegraphics[width=8.00cm]{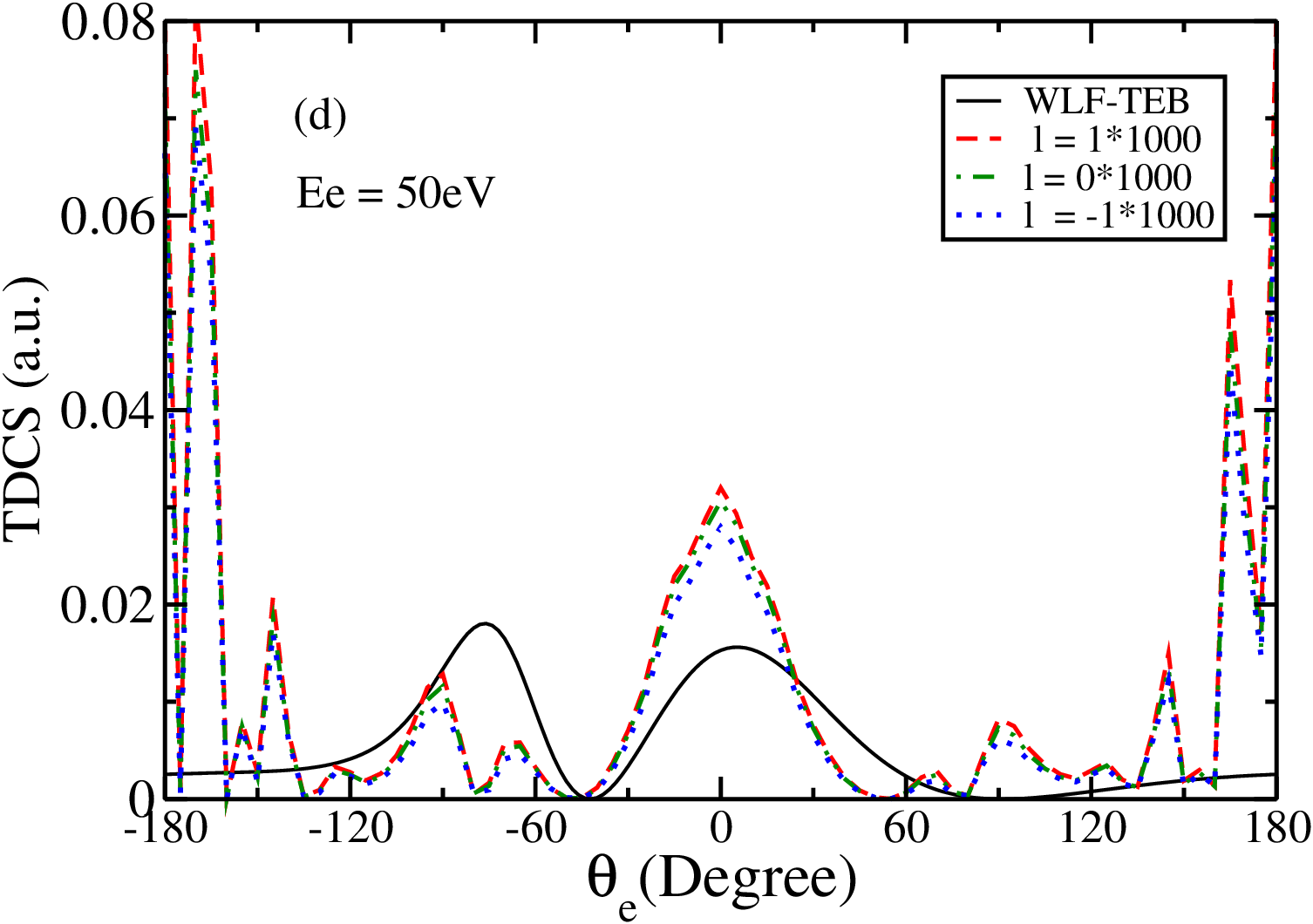}\ & \includegraphics[width=8.00cm]{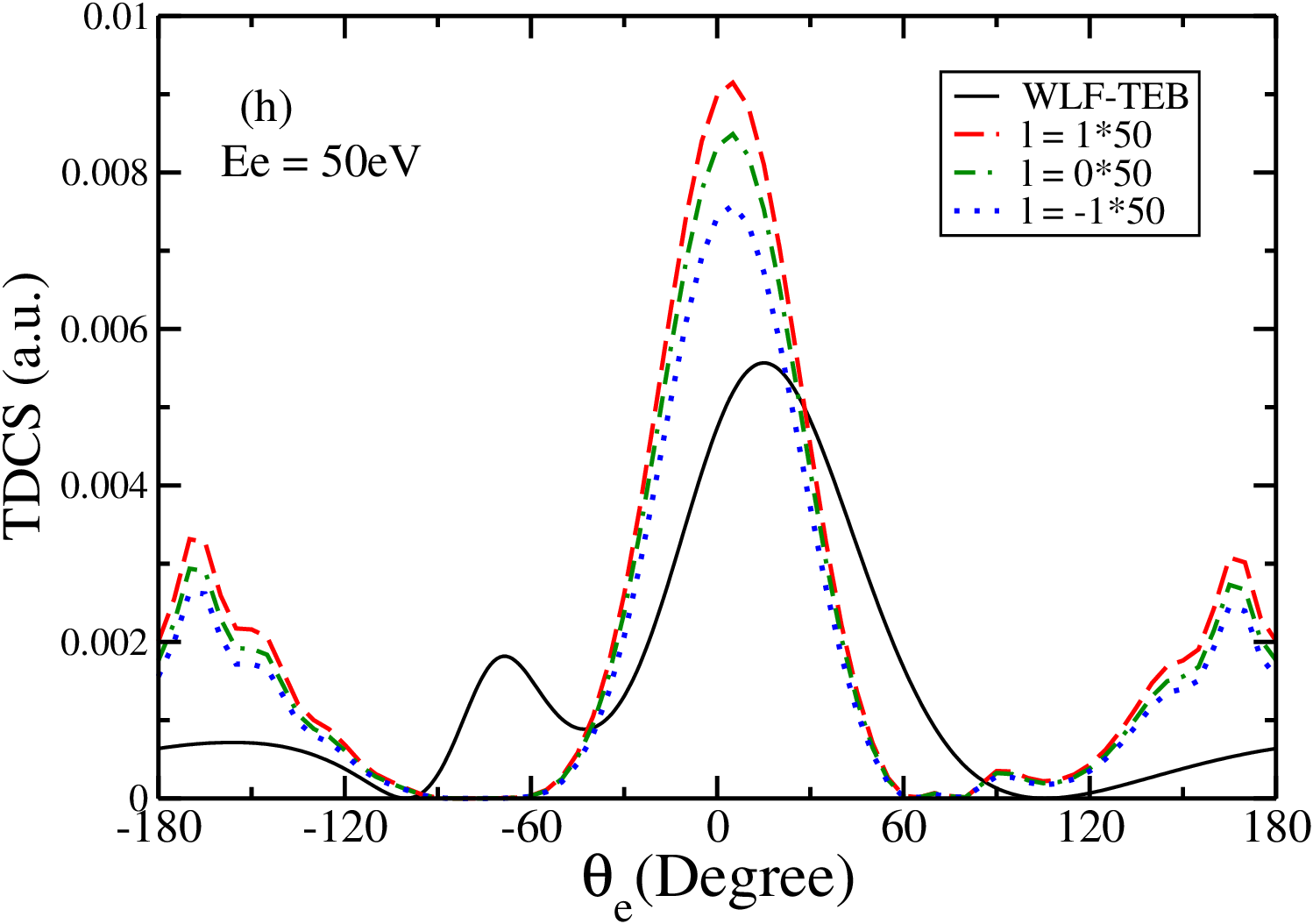}
				\end{tabular}
				
				\caption{TDCS as a function of ejection angle $\theta_e$ for the TEB (e,2e) process on H-atom in the co-planar asymmetric geometry. TEB without laser-field (WLF-TEB) results are represented by a solid line, and dashed, dashed-dotted, and dotted curves represent the LA-TEB for l = 1 (absorption of one photon), l = 0 (no transfer of photon) and l = -1 (emission of one photon). Figures (a)-(d) represent the results for an incident energy 500eV and OAM $m_l$ = 1, while figures (e) - (h) represents the result for OAM $m_l$ = 2. Different ejected electron energies and scaling factors are given in the panel.}   
			\label{fig:9}
			\end{figure*}

			In figures \ref{fig:9} (a-h), we present the results of our calculations of the TDCS for the TEB without laser field (WLF-TEB) and with laser field LA-TEB as a function of the ejected electron angle in coplanar asymmetric geometry with different ejected energy. The calculations are performed in FBA for OAM numbers ($m_l$) 1 and 2 at incident energy 500eV. We observed that the laser field modifies the angular distribution of the TDCS significantly. The order of TDCS magnitude for WLF-TEB is greater than the LA-TEB. For  WLF-TEB calculations, ionization processes are more probable for smaller values of OAM, irrespective of whether OAM is even or odd. But, for LA-TEB, the occurrence of the ionization process is more probable for even values of OAM, i.e., the magnitude of TDCS for even $m_l$ is greater than odd values of $m_l$. As we can see from figures (\ref{fig:9} (a)-(d)) where magnitude of TDCS for $m_l$ = 1 is greater then the $m_l$ = 2  (\ref{fig:9} (e)-(h)) ( black solid line for WLF-TEB ). While the TDCS magnitude is smaller for LA-TEB with $m_l$ = 1  compared to $m_l$ = 2 (see figure \ref{fig:9}(e)-(h)). The scaling factors used facilitate a better comparison of results for different $m_l$ values at varying energies, are indicated in the plot's panel. However, the number of transfer $l$ photons ($l$ = 0, $\pm$ 1) from the laser field to the electron-atom system shows an inverse relation with the TDCS magnitude. On examining the impact of ejected electron energy on the angular profile of TDCS, at $E_e$ = 20eV for  WLF-TEB (see Black solid line in figure \ref{fig:9}(a)), we observed forward and backward peaks along with a shallow recoil peak for $m_l$ = 1 and a dual peak structure (forward and backward peaks) for $m_l$ = 2 (see Black solid line in figure \ref{fig:9}(e)), but LA-TEB only shows two peak structure for both $m_l$ = 1 and 2. A rise in the ejected energy enhances the recoil peaks and reduces the absolute magnitude of TDCS for WLF-TEB incidence ( see black solid curve in figure \ref{fig:9}(a) - \ref{fig:9}(h)). While for LA-TEB with $m_l$ = 2, an increment in the ejected electron energy amplifies the forward peaks and suppresses the backward peaks (see red, green and blue curve in figures \ref{fig:9}(e) - \ref{fig:9}(h)). But contrary to this, we observed only forward peak along many other peaks of varying amplitudes for $m_l$ = 1 at higher ejected energies ($E_e$ = 40eV and $E_e$ = 50eV) (see red, green and blue curves in figure \ref{fig:9}(a) - \ref{fig:9}(e)). We also noticed a symmetry in the angular distribution of TDCS for LA-TEB projectiles with respect to incidence direction. 
			
			\subsection{Angular profile of $(TDCS)_{av}$ for macroscopic H- atomic target}
			\begin{figure*}[htp!]
				\begin{tabular}{ccc}
					\includegraphics[width=6.00cm]{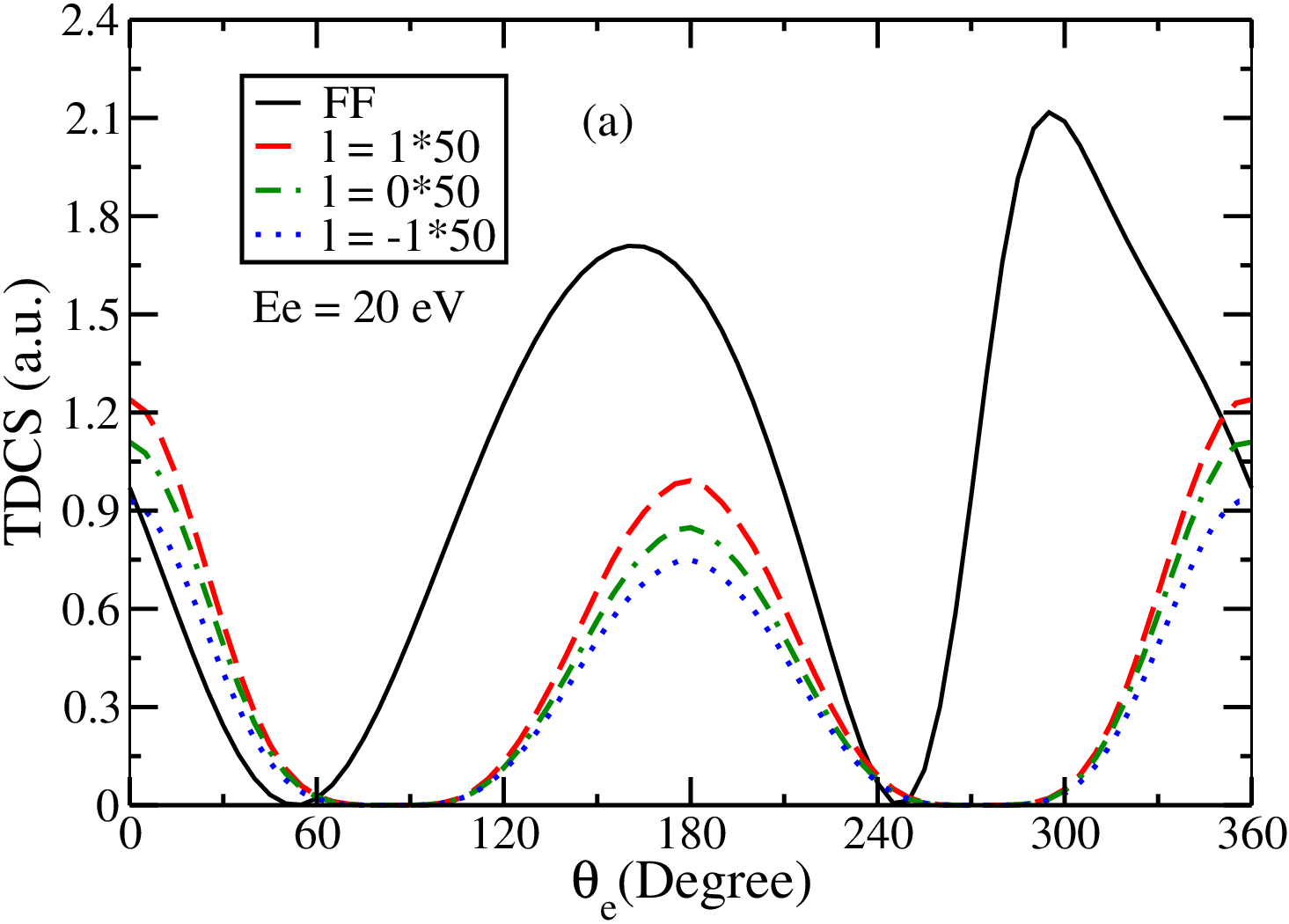}\ & \includegraphics[width=6.00cm]{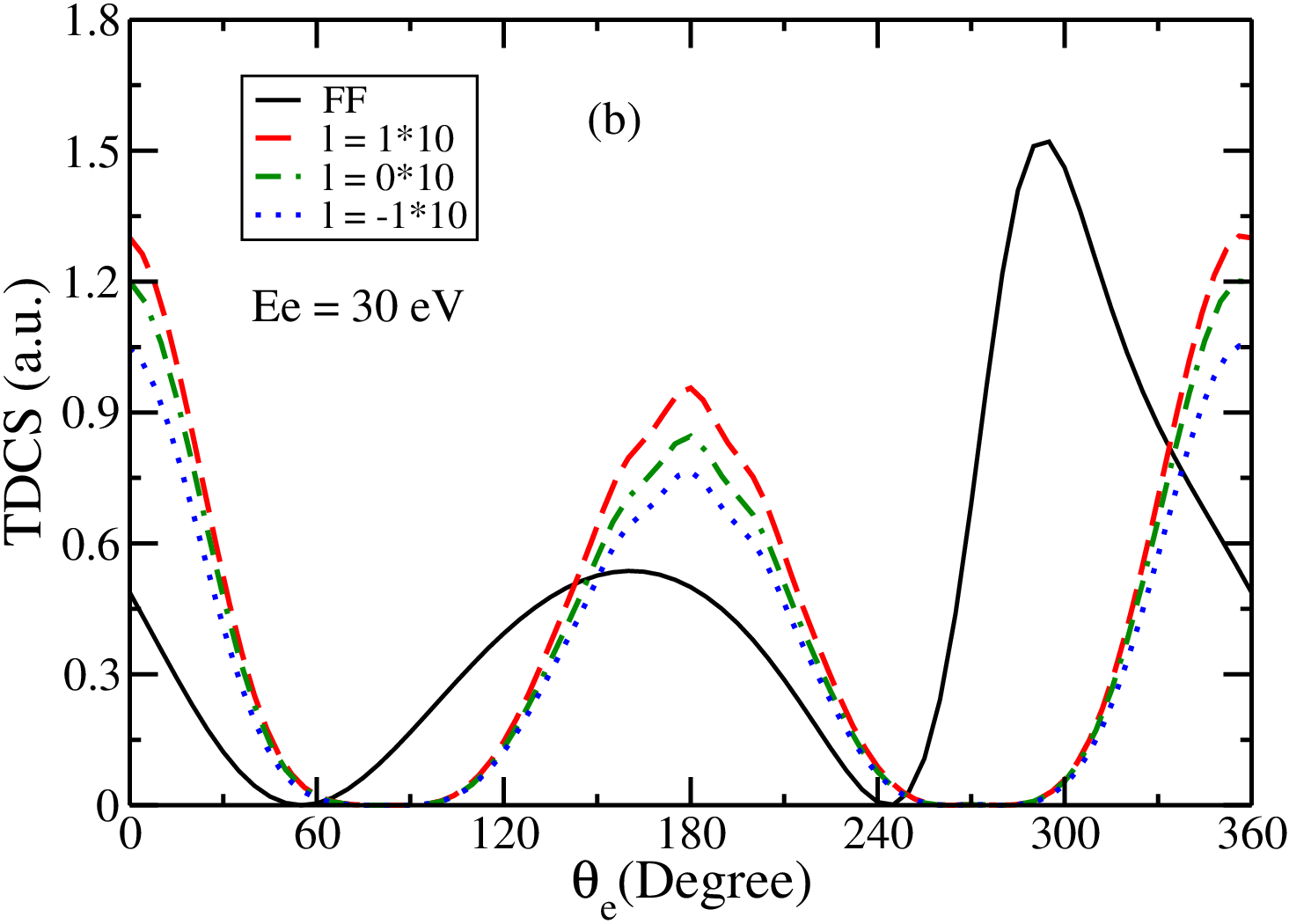}
					& \includegraphics[width=6.00cm]{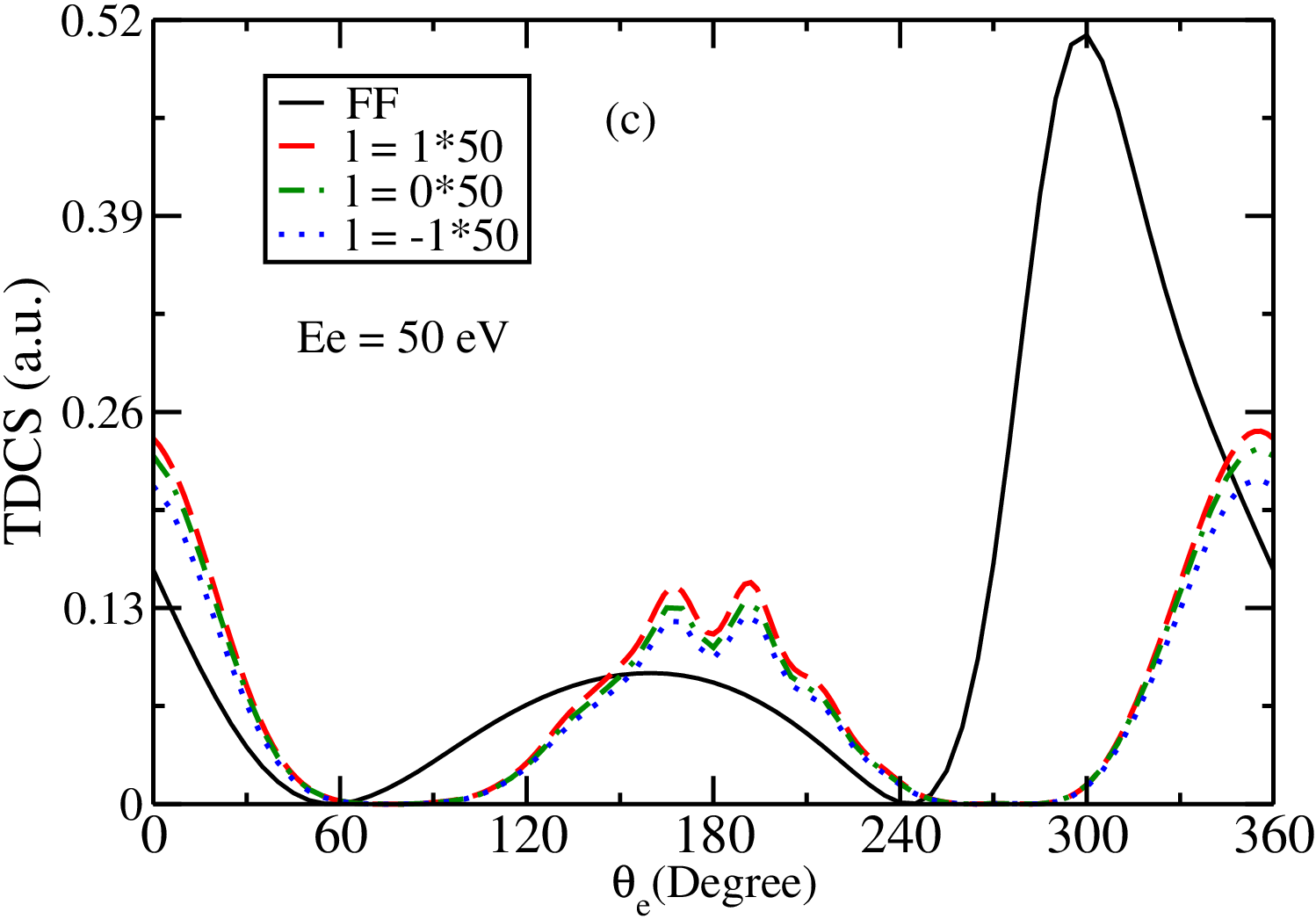}
				\end{tabular}
				\caption{$(TDCS)_{av}$ as a function of ejected electron angle (TDCS integrated over impact parameter b) for the ionization of the H- atom. Different calculations with and without laser field are discribed as:-  WLF-TEB: solid curve, LA-TEB for absorption of one photon (l = 1): dashed curve, for emission of one photon (l = -1): dotted curve and no net transfer of photons (l = 0): dashed-dotted-dashed curve.}  
				\label{fig:10}
			\end{figure*}
			Figure \ref{fig:10} illustrates the outcomes of our calculations for the average of TDCS ($(TDCS)_{av}$), computed using equation (24). As previously stated, the feasibility of conducting an experiment solely focused on on-axis collisions is highly improbable. To obtain a more precise understanding of the probability of ionization caused by TEB in comparison to plane wave projectiles, it becomes necessary to calculate an average across impact parameters. This section discusses the impact of different ejected energies ($E_e$) and the number of photons ($l$) transferred from the laser field. The TDCS of WLF-TEB (TEB without laser field) and LA-TAB (Laser-assisted TEB) decreases when ejected energy increases (see figure \ref{fig:10}(a) - (c)). The TDCS magnitude of LA-TEB (see curves blue, green and red in figure \ref{fig:10}) is smaller than the WLF-TEB (see black solid curve in figure \ref{fig:10})incidence. For TEB projectile without laser-field, we observed dual peak structure, backward peak with a slight shift towards incidence direction, and a peak near $300^\circ$. For LA-TEB projectiles, we observed a symmetry about the incidence direction. We also observed that at $E_e$ = 30eV (see curves blue, green and red in figure \ref{fig:10}(b)), the forward peak is a little distorted and has a kink at $180^\circ$; on further increasing the ejected energy upto 50eV (see curves blue, green and red in figure \ref{fig:10}(c)), we noticed a dip at the same position as $E_e$ = 30eV .i.e. $180^\circ$.
			Considering the effect of the transfer from the laser field, we observed that magnitude of the TDCS for the emission of photons (-$l$) (see blue curve in figure \ref{fig:10}) is smaller then the magnitude of the absorption of photons ($l$) (see red curve in figure \ref{fig:10}) 
			
			\section{Conclusion}
			\label{conc}
			
			In this paper, we have presented the study of the laser-assisted triple differential cross-section for electron impact
			ionization of atomic hydrogen by the plane wave and twisted electron beam in coplanar asymmetric geometry. The effect of the incident and ejected electron energy is studied on TDCS profile for various OAM numbers. The influence of emission and absorption of photons by an electron-atom system in a laser field is also examined. We compared the TDCS of PW, LA-PW, and TEB (without laser field) with the LA-TEB. Unlike the PW, for LA -TEB, peaks are not located in the momentum transfer direction; in fact, we observed a minimum in this direction for LA-TEB. Also, the magnitude of the TDCS projectiles is smaller than the plane wave incidence. \\
			While comparing the TDCS for TEB projectile without laser field (WLF-TEB) and in the presence of laser field (LA-TEB), we observed that TDCS of LA-TEB processes smaller magnitude than WLF-TEB. Further, we found that for WLF-TEB,  ionization probability is higher for smaller values of OAM ($m_l$) and decreases as ($m_l$) increases. But for LA-TEB, ionization probability is higher for even values of OAM, in contrast to the TEB process. Regarding the magnitudes of TDCS, for WLF-TEB, is decreased in the order of $m_l$ as 1, 2, 3, and 4. On the other hand, the magnitude of TDCS for LA-TEB followed the order of $m_l$ as 2, 4, 1, and 3.\\
			We also discuss the TDCS average $(TDCS)_{av}$ for the TEB in the absence of laser field (WLF-TEB) and the presence of laser-field (LA-TEB). We observed that $(TDCS)_{av}$ of LA-TEB is significantly altered as compared to that for the WLF-TEB.\\
			This communication aims to study about the effects of twisted electron beam parameters on the angular profile of TDCS in the presence of a laser field. The present communication is the first attempt to study the laser-assisted (e,2e) process using a twisted electron beam. The current study can be expanded to investigate the impact of additional laser parameters such as frequency, differently polarised light (elliptically polarised, circularly polarised), laser-field strength, etc., in various kinematics such as symmetric coplanar geometry non-coplanar geometry. No experimental or theoretical results of the laser-assisted (e,2e) technique for twisted electron beam exist. However, we hope that this research may inspire additional research on twisted electron beam laser-assisted ionization processes.\\
		\section*{Acknowledgments}
 Authors acknowledge the Birla Institute of Technology and Science Pilani, Pilani Campus, for for financial support. 
			\nocite{apsrev41Control}
			\bibliography{MS_CRSV2}  
		\end{document}